\documentclass[]{tPHM2e}

\begin{document}
\doi{10.1080/14786435.20xx.xxxxxx} \issn{1478-6443}
\issnp{1478-6435}
\jvol{00} \jnum{00} \jyear{2012} %\jmonth{21 December}

\markboth{Taylor \& Francis and I.T. Consultant}{Philosophical
Magazine}

\title{Studies of homogeneous precipitation in very dilute
iron-coper alloys using kinetic Monte Carlo simulations and
statistical theory of nucleation}

\author{V.G. Vaks$^{\rm a,b}$$^{\ast}$
\thanks{$^\ast$Corresponding author. Email:
vaks@mbslab.kiae.ru \vspace{6pt}}, F. Soisson$^{\rm c}$,  and I.A.
Zhuravlev$^{\rm a}$\\
\vspace{6pt} \noindent $^{\rm a}${\em{National Research Center
``Kurchatov Institute'', 123182 Moscow, Russia}}; $^{\rm
b}${\em{Moscow Institute of Physics and Technology (State
University), 117303 Moscow, Russia}}; $^{\rm c}${\em CEA, DEN,
Service de Recherches de M\'etallurgie
Physique,  91191 Gif-sur-Yvette, France}\\
\vspace{6pt}\received{Received \today } }

\maketitle

\begin{abstract}
\noindent{\bf Abstract} -- Kinetics of homogeneous nucleation and
growth of copper precipitates under electron irradiation of
Fe$_{1-x}$Cu$_x$ alloys at concentrations $x$ from 0.06\, to
0.4\,at.\% and temperatures $T$ from 290 to 450$^{\circ}$C  is
studied using the kinetic Monte Carlo (KMC) simulations and the
statistical theory of nucleation (STN). The conventional assumption
about the similarity of mechanisms of precipitation under electron
irradiation and during thermal aging is adopted. The
earlier-developed $ab \ initio$ model of interactions in Fe-Cu
alloys is used for both the KMC simulations and the STN
calculations. Values of the nucleation barrier $F_c$ and the
prefactor $J_0$ in the Zeldovich-Volmer relation for the nucleation
rate $J$ are calculated for a number of concentrations and
temperatures. For the dilute alloys with $x\leq 0.2\%$, the STN and
the KMC results for the nucleation barrier $F_c$ do virtually
coincide with each other, which seems to confirm a high reliability
of the STN for this problem. The STN calculations are also used to
estimate the temperature dependencies of concentrations which
correspond to the homogeneous or the heterogeneous precipitation
limit, $x_l^{hom}(T)$ and $x_l^{het}(T)$, and both dependencies are
found to be rather sharp.

\bigskip

\end{abstract}

%PACS: 05.70.Fh; 05.10.Gg

\section{Introduction}

Studies of copper precipitation in irradiated Fe-Cu-based alloys at
temperatures  $T\sim 290-300^{\circ}$C and low copper concentrations
$x$ between $x$=0.05-0.06\% and $x$=0.2-0.3\% (in atomic percents,
here and below) attract great attention, first of all in connection
with the problem of hardening and embrittlement of the nuclear
reactor pressure vessels (RPVs), see, e. g.
\cite{Akamatsu-95}-\cite{Meslin-10b}. Numerous experiments combined
with some simulations showed that for the usual conditions of
electron irradiation, that is, when the radiation-induced clusters
of vacancies and self-interstitial atoms are not formed and the
heterogeneous precipitation on them is basically absent, SANS and
ATP measurements at $x\lesssim 0.1\%$ typically do not reveal any
copper-rich precipitates \cite{Akamatsu-95,Mathon-97,Radiguet-07},
even though the alloys remain to be strongly supersaturated. The
absence of a significant homogeneous copper precipitation in these
electron-irradiated Fe-Cu-based alloys is commonly explained ``by
the very low volume fraction of probably very small precipitates,...
and ... one can infer that the limit of observable copper
precipitation at 290 $^{\circ}$C is located between $x$=0.09 and
0.08\%\,'' \cite{Mathon-97}, or ``between $x$=0.1 and 0.2\%\,''
\cite{Auger-00}. In particular, cluster dynamics simulations of
homogeneous precipitation at $x$=0.088\%, \hbox{$T=290$$^{\circ}$C}
described in \cite{Radiguet-07} predict ``a very low density
($5\cdot 10^{-16}$ m$^{-3}$) of very big ($R\simeq 30$ nm) copper
precipitates''. Under neutron or ion irradiations when copper
precipitation can occur heterogeneously at  clusters of vacancies or
self-interstitial atoms, such precipitation at $T=290^{\circ}$C was
observed for $x\gtrsim 0.08$ \cite{Radiguet-07,Meslin-10a}, but not
for $x\lesssim 0.06\%$ \cite{Miller-06,Miller-09}.

Therefore, theoretical studies of copper precipitation in very
dilute Fe-Cu-based alloys near the concentration or temperature
limits of such precipitation, $x_l^{hom}(T)$ for the homogeneous
nucleation and  $x_l^{het}(T)$ for the heterogeneous nucleation,
seem to be interesting from both fundamental and applied
standpoints, even for binary Fe-Cu alloys which can be considered as
model alloys for the RPV steels
\cite{Akamatsu-95,Mathon-97,Duparc-02,Radiguet-07}. Such theoretical
studies can be particularly interesting if they can sufficiently
reliably predict the temperature dependencies of these precipitation
limits at elevated $T>290^{\circ}$C for which experimental estimates
are either uncertain or absent.

A consistent $ab \ initio$ model for studies of precipitation
kinetics in Fe-Cu alloys has been recently developed by Soisson and
Fu  \cite{SF-07}, and detailed kinetic Monte Carlo (KMC) simulations
of precipitation in a number of Fe-Cu alloys at different $T$ and
$x$ based on this model yield a good agreement with available
experimental data \cite{SF-07,KVZ-13}. Therefore, this model and KMC
methods developed in \cite{SF-07} seem to be prospective to study
the above-discussed problem of copper precipitation in dilute Fe-Cu
alloys.

We will also use the statistical theory of nucleation and growth of
isolated precipitates (to be abbreviated as STN) developed by
Dobretsov and Vaks \cite{DV-98a,DV-98b}. This theory provides
microscopic expressions for both the nucleation barrier (activation
barrier for the formation of a critical embryo)  $F_c$ and the
prefactor $J_0$ in the phenomenological Zeldovich and Volmer formula
for the nucleation rate $J$ (the number of supercritical embryos
formed in unit volume per unit time) \cite{Lif-Pit-79} valid for the
initial, steady-state nucleation stage of precipitation:
\begin{equation}
J=J_0\exp (-\beta F_c) \label{J-J_0}
\end{equation}
where $\beta =1/T$ is the reciprocal temperature. The STN
corresponds to a number of refinements of earlier models of
nucleation suggested by Cahn-Hilliard \cite{Cahn-Hil-59} and Langer
\cite{Langer-69}. It is based on the generalized Gibbs distribution
approach \cite{Vaks-04} and enables one to quantitatively calculate
values of $F_c$ and $J_0$ for the microscopic alloy model chosen,
particularly in the case of high nucleation barriers, $\beta F_c\gg
1$, which just corresponds to the dilute alloys under consideration.
In description of growth of supercritical embryos we also use the
phenomenological equation of such growth suggested in the classical
theory of nucleation \cite{Lif-Pit-79} and access validity of this
equation by comparison with our KMC simulations.

In all our simulations and calculations we consider only homogeneous
precipitation and accept the conventional assumption (discussed, in
particular, by Mathon et al. \cite{Mathon-97}) that for the moderate
irradiation intensity (i.e. when the ratio of ballistic jump
frequency to thermally activated jump frequency is small), the main
effect of irradiation is to increase the point defect
concentrations, and hence the Fe and Cu diffusion coefficients. It
means that the precipitation kinetics under electron irradiation is
the same as that during thermal aging at the same temperature,
except for an acceleration factor $A_{irr}$ defined by the relation:
\begin{equation}
A_{irr}=D^{irr}_{Cu}/D^{th}_{Cu}\label{D_Cu}
\end{equation}
where $D^{th}_{Cu}$ and $D^{irr}_{Cu}$ are the copper diffusivities
during thermal aging and under irradiation, respectively. As
evolution of microstructure under precipitation is determined by the
diffusion of copper, Eq. (\ref{D_Cu}) implies an analogous relation
between the evolution times $t_{irr}$ and $t_{th}$ under irradiation
and during thermal aging:
\begin{equation}
t_{irr}=t_{th}/A_{irr}. \label{t_irr}
\end{equation}
Then the precipitation kinetics under electron irradiation can be
described by the KMC codes developed by Soisson and Fu \cite{SF-07}
for thermal aging with replacing the thermal aging time $t_{th}$ by
the ``scaled'' time $t_{irr}$ according to Eq. (\ref{t_irr}). This
approach is accepted throughout this work. In more detail, scaling
relation  (\ref{D_Cu}) and estimates of the acceleration factor
$A_{irr}$ are discussed in Sec. 4. At the same time, the main
results and conclusions of this work do not depend on $A_{irr}$
values as this scaling factor is canceled in many important kinetic
characteristics.

In Sec. 2 we present some results of calculations of structure of
critical embryos and parameters $F_c$ and $J_0$ in (\ref{J-J_0}) for
dilute Fe-Cu alloys obtained using  the STN \cite{DV-98a,DV-98b}. In
Sec. 3 we describe our KMC simulations of nucleation and growth of
copper precipitates in dilute Fe-Cu alloys. In Sec. 4 we present an
estimate of the acceleration factor $A_{irr}$ based on the rate
theory models \cite{Sizmann-78} and find this estimate to reasonably
agree with those obtained using the available experimental data
\cite{Mathon-97} and our KMC simulations. In Sec. 5 we use the
phenomenological equation for growth of a supercritical embryo
\cite{Lif-Pit-79} and our KMC simulations to estimate some important
kinetic parameter which enters into the STN expression for the
prefactor $J_0$ in Eq. (\ref{J-J_0}). In Sec. 6 we show that the
nucleation barrier $F_c$ values calculated using the STN agree well
with those estimated in our KMC simulations, particularly at large
$\beta F_c\gg 1$ which correspond to the concentrations and
temperatures near the above-mentioned precipitation limits. In Sec.
7 we use the STN calculations combined with some plausible physical
assumptions to estimate the temperature dependence of both the
homogeneous and the heterogeneous precipitation limit,
$x_l^{hom}(T)$ and $x_l^{het}(T)$, for practically interesting
temperatures $T$ between 290 and 450$^{\circ}$C. Our main
conclusions are summarized in \hbox{Sec. 8.}

\section{Properties of critical embryos in dilute Fe-Cu
alloys calculated using the statistical theory of nucleation}

In this section we use the STN described in Refs.
\cite{DV-98a,DV-98b} (to be referred to as I and II) to calculate
some thermodynamic, structural  and kinetic characteristics of
critical embryos. For these calculations we use the $ab$ $initio$
model of Fe-Cu alloys developed by Soisson and Fu and described in
detail in Ref. \cite{SF-07}. Here we only note that this model uses
the following values of the binding energy between two copper atoms
and between a copper atom and a vacancy, $E^{bn}_{\rm CuCu}$ and
$E^{bn}_{v{\rm Cu}}$, for the $n$-th neighbors (in eV):
\begin{eqnarray}
&&E^{b1}_{\rm CuCu}=0.121-0.182T,\qquad E^{b2}_{\rm
CuCu}=0.021-0.091T,\nonumber\\
&&E^{b1}_{v{\rm Cu}}=0.126,\qquad E^{b2}_{v{\rm Cu}}=0.139.
\label{E^b-SF}
\end{eqnarray}
The high values of  $E^{bn}_{\rm CuCu}$ in Eqs. (\ref{E^b-SF})
correspond to strong thermodynamic driving forces for precipitation,
while a strong attraction between vacancy and copper atoms leads to
the strong  trapping of vacancies by copper precipitates discussed
in detail in \cite{SF-07}.

First we consider the structural and thermodynamic characteristics
of embryos which are described in the STN in terms of the occupation
number \,$n_i$ for each lattice site $i$.\,  The operator $n_i$ is
unity when a copper atom is at site $i$ and zero otherwise, while
probabilities of various distributions $\{n_i\}$ are described by
the distribution function $P\{n_i\}$ given by Eq. (I-3):
\begin{equation}
P\{n_i\}=\exp [\beta(\Omega+\mu\sum_in_i-H)]\,.\label{P}
\end{equation}
Here $H$ is the configurational Hamiltonian supposed to be pairwise,
$\mu$ is the chemical potential, and the grand canonical potential
$\Omega$ is determined by the normalization condition:
\begin{equation}
H=\sum_{i>j}v_{ij}n_in_j,\qquad \Omega =-T\ln {\rm Tr}\exp
[\beta(\sum_i\lambda_in_i-H)]\label{H_Omega}
\end{equation}
where $v_{ij}$ are the configurational interactions, while symbol
``Tr'' means summation over all configurations $\{n_i\}$. The free
energy $F$ of an alloy  with embryo is determined by relations
\begin{equation}
F=\Omega+\mu\sum_ic_i\, ,\qquad \mu=\partial F/\partial c_i={\rm
constant}\label{F-mu}
\end{equation}
where $c_i=\langle n_i \rangle = {\rm Tr}(n_iP)$ is the mean
occupation number, or the local concentration. The free energy
$F=F\{c_i\}$ in (\ref{F-mu}) was calculated using several
statistical approximations described in detail in I and II:
mean-field, ``mean-field with fluctuations'', pair-cluster, and
``pair-cluster-with fluctuations''. The latter was found to be more
consistent and accurate than others, thus below we use only the
``pair-cluster-with fluctuations'' approximation.

In tables 1 and 2 and Fig. 1 we present some characteristics of
critical embryos. To illustrate both concentration and temperature
dependencies of these characteristics, we consider two series of
Fe$_{1-x}$Cu$_x$ alloys: those with the same temperature
$T=290^{\circ}$C but different concentrations $x$ varying from
$0.06$ to $0.4$\% (Table 1), and those with the same concentration
$x=0.2\%$ but different temperatures $T$ varying from $290$  to
390$^{\circ}$C (Table 2). In these tables, $N_c$ is the total number
of copper atoms within the embryo, and $R_c$ is its effective radius
defined as that of the sphere having the same volume as $N_c$ copper
atoms in the BCC lattice of $\alpha$-iron with the lattice constant
$a_0$=0.288 nm:
\begin{equation}
R_c=a_0(3N_c/8\pi)^{1/3}=0.142\,N_c^{1/3}\,{\rm nm}. \label{R_c}
\end{equation}
Note that this critical radius (as well as the precipitate radius
$R$ in Eq. (\ref{R-N}) below) characterizes the total number of
atoms  but not the ``geometrical'' size of precipitate, in
particular, not its mean squared radius $\langle r_i^2\rangle^{1/2}$
defined by Eq. (II-10). Therefore, for small and ``loose''
precipitates, such as those shown by lines B and C in Fig. 1, these
radii can notably differ from ``geometrical'' ones. However,
quantities $R_c$ and $R$ in (\ref{R_c}) and (\ref{R-N}) are
convenient to describe the precipitation kinetics.

Quantities $\Delta\Omega_0$ and $\Delta\Omega_1$ in tables 1 and 2
are the zero-order and the first-order terms in fluctuative
contributions to the total nucleation barrier
$F_c=\Delta\Omega_0+\Delta\Omega_1$ (defined as the difference
between grand canonical potentials of two alloy states, that with
the embryo and that without the embryo). These two terms correspond
to the iterative treatment of some exact relation of thermodynamics
of nonuniform systems given by Eq. (I-9) which relates the free
energy $F=F\{c_i\}$ to the correlator $K_{ij}$ of fluctuations of
site occupations $n_i$: $K_{ij}=\langle (n_i-c_i)(n_j-c_j)\rangle$.
As discussed in II and below, the accuracy of such iterative
treatment of fluctuations for dilute alloys considered is usually
rather high. Fig. 1 illustrates variations of structure of critical
embryos with concentration or temperature.

The results presented in tables 1, 2 and Fig.1 clearly illustrate
very sharp variations of both structure and thermodynamics of
critical embryos with concentration $x$ and temperature $T$. In
particular, at $T=290^{\circ}$C, the decrease of concentration from
$x=0.4$ to $x=0.06\%$ leads to the increase of the embryo size $N_c$
and the reduced nucleation barrier $\beta F_c$ by about three times:
from $N_c\simeq 8$ to $N_c\simeq 25$, and from $\beta F_c\simeq 12$
to $\beta F_c\simeq 35$. According to the Zeldovich-Volmer relation
(\ref{J-J_0}), the latter implies decreasing the nucleation rate by
about ten orders of magnitude (variation with $x$  of the prefactor
$J_0$ in  (\ref{J-J_0}) will be shown to be negligible). Table 2
illustrates a similar sharp increase of $N_c$ and $\beta F_c$ under
elevating temperature between 290 and $390^{\circ}$C. Fig. 1 shows
that this sharp rise of sizes and nucleation barriers is accompanied
by notable changes in the structure of the embryo: its boundary at
high $\beta F_c\gtrsim 30$ is much less diffuse than that at
moderate $\beta F_c\lesssim 20$.

Discussing the concentration profiles shown in Fig. 1 we note that
these profiles (as well as those shown in Fig. 1 in \cite{KVZ-13}
and Fig. 2 in \cite{KSSV-11}) usually correspond to rather diffuse
interfaces. At the same time, KMC simulations show that only very
small clusters can have a diffuse interface, while for $R\gtrsim
0.3$ nm (that is, $N\gtrsim 10$ copper atoms) one observes almost
``pure'' copper clusters with a sharp interface, see, e.g., Fig. 8
in \cite{SF-07}. However, one should take into account that the
profiles shown in Fig. 1 (and other similar figures) correspond to
statistical averaging over all orientations of a cluster which is
typically rather anisotropic, and this averaging leads to diffuse
interfaces; it is illustrated, in particular, by Fig. 8 in
\cite{SF-07}. Let us also note that in phenomenological treatments
of critical embryos based on the Cahn-Hilliard continuum approach
\cite{Nagano-06,Philippe-11}, the embryo interfaces are usually
notably more diffuse than those obtained in the microscopic STN
used. In particular,  rather sharp interfaces shown by curves A and
D in Fig. 1 can hardly be obtained in the phenomenological
treatments.

Let us now discuss the prefactor $J_0$ in Eq. (\ref{J-J_0}). In the
STN, it is given by Eq. (II-3):
\begin{equation}
J_0=\left(\beta |\gamma_0|/2\pi\right)^{1/2}{\cal N} D_{\bf R}({\bf
u})D_{aa}.\label{J_0}
\end{equation}
Here the first three factors (discussed in detail in I) have the
``thermodynamic'' origin. They describe dependencies of the free
energy and the distribution function of the embryo on the
``critical'' variable $a$ characterizing its size, and on three
variables $\bf u$ characterizing tits position $\bf R$. The last
factor $D_{aa}$ is the generalized diffusivity  in the $c_i$-space
that corresponds to an increase of $a$, i. e. to growth of the
embryo. This factor is defined by Eq. (I-55) which expresses
$D_{aa}$ as a certain linear combination of generalized mobilities
$M_{ij}$ which describe the temporal evolution of mean occupations
$c_i$ via Eq. (I-50):
\begin{equation}
dc_i/dt=\sum_j[M_{ij}-\delta_{ij}\sum_kM_{ik}]\beta\partial
F/\partial c_j.\label{dc_i/dt}
\end{equation}
The $M_{ij}$ values can be calculated using some microscopic models,
for example, those employed in II. In particular, for the ideal
solution (or in the dilute alloy limit which corresponds to small
$c_i$), the mobility $M_{ij}=M_{ij}^0$ is given by Eq. (II-8) with
$u_{ik}=0$:
\begin{equation}
M_{ij}^0=\delta_{ij,nn}\gamma_{nn}^0[c_i(1-c_i)c_j(1-c_j)]^{1/2}.\label{M_ij^0}
\end{equation}
Here $\delta_{ij,nn}$ is the Kroneker symbol equal to unity when
sites $i$ and $j$ are the nearest neighbors and zero otherwise,
while quantity $\gamma_{nn}^0$ has the meaning of the mean rate of
exchanges between neighboring copper and iron atoms. It is related
to the copper diffusivity $D_0$ in a pure iron by the following
relation (see, e. g., Eqs. (72) and (74) in \cite{VZh-12}):
\begin{equation}
\gamma_{nn}^0=D_0/a_0^2\label{gamma_nn^0}
\end{equation}
where $a_0$ is the BCC iron lattice constant used in Eq.
(\ref{R_c}).

At the same time, mobilities $M_{ij}$ can  be considered as
phenomenological parameters in the general Onsager-type equations
(\ref{dc_i/dt}), and for weakly nonuniform alloys under
consideration they can be supposed to be nonzero only for
neighboring sites $i$ and $j$, just as in Eq. (\ref{M_ij^0}). For
our problem, the phenomenological model of $M_{ij}$ can be
considered as adequate if it properly describes the diffusion of
solute (copper) atoms in processes of nucleation and growth of
embryos. In Sec. 5 we show that growth of a supercritical embryo in
dilute Fe-Cu alloys can be described well using some generalization
of the phenomenological diffusion equation suggested in the
classical theory of nucleation \cite{Lif-Pit-79}  given by Eq.
(\ref{dR/dt}) below. This generalized diffusion equation differs
from those for ideal or dilute solutions mainly by the value of the
effective diffusivity $D_{eff}$ which takes into account  the
acceleration of precipitation kinetics due to the high mobility of
small copper clusters noted in Refs. \cite{SF-07,Jourdan-10} and
discussed below. The $D_{eff}$ values for dilute Fe-Cu alloys
considered are estimated in Sec. 5 using kinetic Monte Carlo
simulations.

Therefore, we suggest that the adequate phenomenological description
of precipitation in Fe-Cu alloys under consideration  can be
obtained if we assume for mobilities $M_{ij}$ in (\ref{dc_i/dt}) the
same simplest form as that in Eq. (\ref{M_ij^0}),
\begin{equation}
M_{ij}=\delta_{ij,nn}\gamma_{nn}[c_i(1-c_i)c_j(1-c_j)]^{1/2}\,,
\label{M_ij}
\end{equation}
but assume the effective exchange rate $\gamma_{nn}$ in (\ref{M_ij})
to be proportional  not to the diffusivity $D_0$ in a pure iron as
in Eq.  (\ref{gamma_nn^0}),  but to the phenomenological effective
diffusivity $D_{eff}$ mentioned above:
\begin{equation}
\gamma_{nn}=D_{eff}/a_0^2.\label{gamma_nn}
\end{equation}
Validity of this assumption will be checked by comparison with the
KMC simulation results presented in Secs. 5 and 6.

In Table 3 we present values of various factors in the expression
(\ref{J_0}) for the prefactor $J_0$ in Eq. (\ref{J-J_0}) for some
alloys Fe-Cu. Quantity $\gamma_0$ in this table is the derivative of
the free energy $F$ with respect to the effective embryo size $a$:
$\gamma_0=\partial^2F/\partial a^2$.  As the critical embryo
corresponds to the saddle-point of $F$ in the $c_i$-space with
respect to this size (I, II), $\gamma_0$ values in table 3 are
negative. ${\cal N}$ is the normalizing constant in the embryo size
distribution function; $D_{aa}$ is the generalized diffusivity
discussed above, and $D_{\bf R}({\bf u})$ is some geometrical factor
which for a large embryo is proportional to its surface. For
quantities $D_{aa}$, $D_{\bf R}({\bf u})$, and the total prefactor
$J_0$ we present their ``reduced'', dimensionless values:
$D_{aa}/\gamma_{nn}$, $v_aD_{\bf R}({\bf u})$, and
$\tilde{J_0}=v_aJ_0/\gamma_{nn}$. Note that the embryo size $N_c$
decreases with $x$ and  increases with  $T$, as tables 1 and 2 show.

Table 3 shows that both concentration and temperature dependencies
of quantities ${\cal N}$ and $D_{aa}/\gamma_{nn}$ are rather weak.
On the contrary, the geometrical factor $D_{\bf R}({\bf u})$ notably
increases with the embryo size $N_c$, while $\beta|\gamma_0|$
somewhat decreases with $N_c$. The total reduced prefactor
$\tilde{J_0}= v_aJ_0/\gamma_{nn}$ notably increases with $N_c$, by
about 20 or 5 times for the total $x$ or $T$ intervals shown in
table 3. However, in the total nucleation rate $J$ given by Eq.
(\ref{J-J_0}), these variations of $J_0$ are negligible as compared
to the above-mentioned huge changes of the activation factor
$\exp\,(-\beta F_c)$ for these intervals of $x$ or $T$.

Now we note that a very sharp decrease of this activation factor
with $x$ between $x\sim 0.2$ and $x\sim 0.08\%$ seen in table 1
evidently correlates with the above-mentioned experimental estimates
of the homogeneous precipitation limit at $T=290^{\circ}$C given in
Refs. \cite{Mathon-97,Auger-00,Radiguet-07}. It seems natural to
suggest that the position  $x_l^{hom}$ of this limit is mainly
determined by the activation factor $\exp (-\beta F_c)$, or by the
reduced nucleation barrier $\beta F_c$ which at $x=x_l^{hom}$
reaches a certain high value $(\beta F_c)_l$. For example, if we
accept for the experimental $x_l^{hom}$=$x_l$ at $290^{\circ}{\rm
C}$ the estimate from Ref. \cite{Mathon-97}, $x_l^{exp}\sim
0.08$-$0.09\%$, we have: $(\beta F_c)_l\simeq 28$, while taking this
estimate from Ref. \cite{Auger-00}, \,$x_l^{exp}\sim 0.15\%$, we
obtain: $(\beta F_c)_l\simeq 23$. In more detail, estimates of the
precipitation limits $x_l$ are discussed below in Sec. 6.

\section{Kinetic Monte Carlo simulations of nucleation and growth
of precipitates in dilute Fe-Cu alloys}

As mentioned in Sec. 1, our KMC simulations use the ``scaling''
assumptions (\ref{D_Cu}) and (\ref{t_irr}) which enable us to relate
the homogeneous precipitation in irradiated alloys to that during
thermal aging for which the  KMC codes developed by Soisson and Fu
\cite{SF-07} can be employed. In these simulations we consider the
following Fe$_{1-x}$Cu$_x$ alloys: (A) those  at $T=290^{\circ}$ C
and $x$ equal to 0.088, 0.116, 0.144, 0.177, 0.2, 0.3 and 0.4\%, and
(B) those at $x=0.2\%$ and temperatures $T$ equal to 290, 305, 320,
335, 350, 365, 380 and 390$^{\circ}$C. We use simulation volume
$V_s=L^3$ with periodic boundary conditions and, to check the
statistical reliability of results, we usually employ several
different values of the simulation size $L$:
\begin{equation}
L_1=128a_0,\quad L_2=160a_0, \quad L_3=192a_0.\label{L_n}
\end{equation}

Before to describe the results, we mention a characteristic feature
of such simulations for some low concentrations $x$ or high
temperatures $T$: at such $x$ and $T$,  we observe only one
precipitate within simulation volume, that is, the total number
$N_p$ of precipitates within simulation box is unity (while at
further lowering $x$ or elevating $T$ we observe no precipitates for
the simulation time). In our simulations it was the case for the
series (A) alloys with $T=290^{\circ}$C at $x\leq 0.144\%$, and for
the series (B) alloys with $x=0.2\%$, at $T\geq 335^{\circ}$C. In
all such cases, the relation $N_p=1$ preserves under increasing
simulation volume (replacing $L_1$ by $L_2$ or $L_3$),  though the
incubation time $t_{inc}$ (that which precedes to the formation of
the embryo \cite{SM-00}) somewhat decreases when $V_s$ increases.

The presence of changes of physical characteristics (such as the
precipitate density $d_p=N_p/V_s$ or  the incubation time $t_{inc}$)
under variations of simulation volume evidently indicates on the
statistical unreliability of simulation results for these
characteristics. At the same time, the growth of a precipitate after
its formation seems to be described by such simulations quite
properly. Therefore, even though the simulations with $N_p=1$ can
not be used to study the evolution of precipitate density $d_p(t)$,
they will be widely used for studies of precipitate growth described
in Sec. 5.

When the number of precipitates within simulation box significantly
exceeds unity, the simulations become statistically reliable and can
be used to study the precipitate density $d_p(t)$. In figures 2 and
3 we show the dependencies $d_p(t)$ for all our simulations for
which the final number of precipitates within simulation box is not
too small: $N_p\geq 3$. The simulation time on abscissa axis is
given in the ``thermal aging'' values $t_{th}$ which in our model
are related to the observed time $t_{irr}$ by Eq. (\ref{t_irr}). The
acceleration factor $A_{irr}$ in this relation and comparison with
the available data \cite{Mathon-97} about precipitation in an
Fe-0.3\%Cu  alloy under electron irradiation are discussed below in
Sec. 4. In the lower part of frames 2(e) and 2(f) we also show
temporal dependencies of the mean precipitate radius $R_m$ for two
simulations with $x=0.3\%$. For other our simulations with
$N_p\gtrsim 10$ these dependencies are similar and describe usually
a smooth increase of $R_m$ by about 20-30\%.

The dependencies $d_p(t_{th})$ presented in Figs. 2 and 3 are
approximately linear. Therefore, if we define the nucleation rate
$J=J_{\rm KMC}$  for these simulations as the ratio
\begin{equation}
J_{\rm KMC}=d_p(\tau)/\tau\label{J_KMC}
\end{equation}
where  $\tau=(t_{th}-t_{th}^{inc})$ is the evolution time counted
off the incubation time $t_{th}^{inc}$, this nucleation rate for
each simulation remains approximately constant, particularly for the
simulations with not small  $N_p$ where fluctuations are not too
strong. Hence these our simulations seem to describe mainly the
first stage of precipitation, the steady-state nucleation, for which
the constant value of the nucleation rate $J$ is characteristic.

The dashed line in each frame of Figs. 2 and 3 shows our estimate of
the $J_{\rm KMC}$ value in Eq. (\ref{J_KMC}) assuming it to be
constant. For this estimate we tried to use the maximum broad
interval of time $t_{th}$ for which the temporal dependence
$d_p(t_{th})$ is close to linear. In spite of the evident scatter of
KMC results presented in Figs 2 and 3, particularly for simulations
with not large $N_p$, the linear temporal dependence
$d_p(\tau_{th})$ can usually be followed rather clearly . Therefore,
our estimates of $J_{\rm KMC}$ values in Eq. (\ref{J_KMC}) seem to
be sufficiently definite.

At the same time, the Zeldovich-Volmer relation (\ref{J-J_0}) is
valid just for the steady-state nucleation stage. Therefore, we can
use our estimates of $J_{\rm KMC}$ to estimate the parameters $F_c$
and $J_0$ in  (\ref{J-J_0}) putting:
\begin{equation}
J_0\exp (-\beta F_c)=J_{\rm KMC}.\label{J_KMC-F_c}
\end{equation}
As discussed in Sec. 2, variations of the activation factor $\exp
(-\beta F_c)$ with concentration or temperature affect the total
nucleation rate $J$ much stronger than those of the prefactor $J_0$.
Therefore, for the given  $J_{\rm KMC}$, the activation factor $\exp
(-\beta F_c)$ can be determined from Eq. (\ref{J_KMC-F_c}) rather
accurately even if the prefactor $J_0$ is estimated not too
precisely, in particular, when $J_0$ is estimated from the STN
calculations described in Sec. 2 which include model assumptions
(\ref{M_ij}) and (\ref{gamma_nn}).

Substituting for the prefactor $J_0$ its expression via quantities
$\tilde{J}_0$, $v_a$ and $\gamma_{nn}$ given in the last line of
table 3, and using  Eq. (\ref{gamma_nn}) which relates $\gamma_{nn}$
to the effective diffusivity $D_{eff}$ mentioned in Sec. 2, we can
re-write relation (\ref{J_KMC-F_c}) as the equation for the reduced
nucleation barrier $\beta F_c$:
\begin{equation}
(\beta F_c)_{\rm KMC}=\ln \Big(2\,D_{eff}\tilde{J}_0/a_0^5 J_{\rm
KMC}\Big)\label{F_c-KMC}
\end{equation}
where  index ``KMC'' in the left-hand side indicates that this
expression for $\beta F_c$ is based on the KMC simulations.

Values of the reduced prefactor $\tilde{J}_0$ can be taken from our
STN calculations illustrated by the last line of table 3. Hence to
find the reduced nucleation barrier $\beta F_c$ from Eq.
(\ref{F_c-KMC}), we need the effective diffusivity $D_{eff}$.
Estimates of  $D_{eff}$ based on our KMC simulations of growth of
precipitates are described below in Sec. 5.

\section{Estimate of acceleration precipitation factor $A_{irr}$ for
electron irradiation of dilute iron-copper alloys}

The concentrations of point defects (vacancies and interstitials)
under permanent irradiation may frequently exceed their equilibrium
values by several orders of magnitudes \cite{Sizmann-78}. In such
conditions diffusive phase transformations, such as precipitation or
ordering, are strongly accelerated. For the case of electron
irradiation of Fe-Cu alloys this has been observed, in particular,
by Mathon et al. \cite{Mathon-97} and by Radiguet et al.
\cite{Radiguet-07}. If other irradiation effects (such as ballistic
mixing or radiation induced segregation) can be neglected, it is
natural to expect that the precipitation kinetics under irradiation
is the same as that during thermal aging at the same temperature,
except for the strong acceleration of this kinetics described by the
factor $A_{irr}$ in  Eqs. (\ref{D_Cu}) and (\ref{t_irr}). In this
section we estimate this acceleration  factor for dilute Fe-Cu
alloys under typical conditions of electron irradiation
\cite{Mathon-97} and compare this estimate with both the
experimental and KMC simulation results.

For dilute alloys under consideration, the copper diffusivity under
thermal aging  in  Eq. (\ref{D_Cu}) is given by the conventional
expression \cite{LeClaire-78}:
\begin{equation}
D_{Cu}^{th}  = \alpha _v c_v^{eq} D_v\label{D_Cu^th}.
\end{equation}
Here $c_v^{eq}$ is the equilibrium vacancy concentration which is
expressed via the enthalpy and entropy of  vacancy formation,
$H_v^{for}$ and $S_v^{for}$, as: $c_v^{eq}=\exp(S_v^{for}-\beta
H_v^{for})$, and $D_v$ is the vacancy diffusion coefficient in a
pure iron which is related to the vacancy migration enthalpy  and
entropy, $ H_v^{mig}$ and $S_v^{mig}$,  as follows:
\begin{equation}
D_v=a_0^2 \nu _0 \exp(S_v^{mig}-\beta H_v^{mig})\label{D_v}
\end{equation}
where $a_0$ is the BCC lattice constant and $\nu_0$ is the attempt
frequency supposed to have the order of the Debye frequency. The
coefficient $\alpha_v$ in (\ref{D_Cu^th}) describes vacancy-solute
correlations, and it is commonly written as \cite{LeClaire-78}:
\begin{equation}
\alpha_v  = f_2 \exp \left({\beta G_{Cu-v}^{bin}
}\right)\label{alpha_v}
\end{equation}
where $f_2$ is the impurity correlation factor,  and
$G_{Cu-v}^{bin}$ is  the copper-vacancy binding energy. These
parameters may be obtained from experimental measurements, or from
{\it ab initio} calculations \cite{SF-07}. Similarly, under
irradiation one may write:
\begin{equation}
D_{Cu}^{irr}  = \alpha _v c_v^{irr} D_v  + \alpha _i c_i^{irr}
D_i\label{D_irr} ,
\end{equation}
where $c_v^{irr}$ and $c_i^{irr}$ are the vacancy and the
interstitial concentrations under irradiation.

It seems natural to assume that the copper-vacancy correlation
effects and binding are not modified by irradiation. Detailed
information on the interstitial diffusion (needed for calculations
of $\alpha _i$ and $D_i$) is usually more difficult to obtain.
However, when the point defect concentrations reach their
steady-state value, one can show that $c_v^{irr} D_v =c_i^{irr} D_i
$ \cite{Sizmann-78}. If $\alpha_v$ and $\alpha _i$ are of the same
order of magnitude, one finally gets $D_{Cu}^{irr} \simeq 2\alpha _v
c_v^{irr} D_v$, which corresponds to the following acceleration
factor in (\ref{D_Cu}) and (\ref{t_irr}):
\begin{equation}
A_{irr} = 2\,(c_v^{irr}/ c_v^{eq})\,.\label{A_irr-FS}
\end{equation}

To compare our KMC simulations during thermal aging and the
experimental kinetics of precipitation under irradiation, the key
point is therefore to get a reliable estimation of the vacancy point
defect concentrations under irradiation. According to the rate
theory models \cite{Sizmann-78}, if point defects created by
irradiation disappear by mutual recombination or by annihilation at
dislocations (which are assumed to be the dominant point defect
sinks), the evolution of their concentrations in a pure metal is
described by the following equations:
\begin{eqnarray}
&&{dc_v\over dt} = K - Rc_i c_v  - \rho_d D_v (c_v  - c_v^{eq} )\nonumber\\
&&{dc_i\over dt} = K - Rc_i c_v  - \rho_d D_i (c_i  - c_i^{eq} )
\simeq K - Rc_i c_v  - \rho_d D_i c_i\,.\label{eq:dcdt}
\end{eqnarray}
Here $K$ is the rate of formation of Frenkel pairs under electron
irradiation (in dpa/s), $\rho_d$ is the dislocation density, and $R$
is the vacancy-interstitial recombination rate:
\begin{equation}
R = 4\pi r_{vi}(D_v  + D_i )/v_a\label{R-recombin}
\end{equation}
where $r_{vi}$ is the recombination radius, and $v_a=a_0^3/2$ is the
atomic volume.

For pure iron (for which lattice constant is $a_0$=0.288 nm), most
of the point defect properties have been estimated by {\it ab
initio} calculations. In particular, for the vacancy formation
parameters we have: $H_v^{for}  = 2.18\;{\rm{eV}}$ \cite{SF-07} and
$S_v^{for}  = 4.08$ \cite{Nichols-78}, which gives: $c_v^{eq} \simeq
1.8 \times 10^{ - 18} $ at 290$^{\circ}$C. The migration enthalpies
are: $H_v^{mig}  = 0.68$~eV~\cite{SF-07} and $H_i^{mig}  =
0.34$~eV~\cite{Lucas-09}. The Debye frequency is $\nu _0 = 10^{13}
{\rm{s}}^{{\rm{-1}}}$, and with these parameters a vacancy migration
entropy $S_v^{mig}  \simeq 2.2$ is required to get the experimental
pre-exponential factor for self-diffusion of iron (see
Refs.~\cite{SF-07, Fu-04} for details). Lacking both experimental
and {\it ab initio} estimations, we will assume the same value for
the migration entropy of the interstitials.

The irradiation conditions of ref. \cite{Mathon-97} are: $K = 2
\times 10^{-9}$ \, dpa/s at 290$^{\circ}$C, with dislocation
densities between $\rho _d=10^8\; \rm cm^{-2}$ and $\rho
_d=10^{11}\; \rm cm^{-2}$. The evolution of point defect
concentrations, obtained by numerical integration of
Eq.~(\ref{eq:dcdt}) with these two dislocations densities, is given
on Fig. 4. In both cases, the steady-state values are reached very
rapidly (after less than $10^{-1}$~s), with vacancy concentrations
$c_v^{irr}  \simeq 3.5 \times 10^{-11}$  and $3.6 \times 10^{ - 14}
$. The acceleration factor is therefore $A_{irr} \simeq 4 \times
10^4 $ for the highest dislocation density and $A_{irr} \simeq 3.9
\times 10^7 $ for the lowest.

In Ref. \cite{Mathon-97}, a precipitate density of $0.9\times
10^{23} \rm m^{-3}$ is observed under electron irradiation of
Fe-0.3\%Cu alloy after approximately $8.3\times10^4$ s. Our KMC
simulations of precipitation during thermal aging shown in Fig. 2(f)
predict a similar density after $1.5\times 10^9$ s. This corresponds
to an acceleration by a factor $A_{irr}\sim 1.8\times 10^4$ under
irradiation. Considering the strong approximations used in the rate
theory model, and the uncertainties on the point defect properties,
the agreement between our estimate of $A_{irr}$ and the combination
of experimental and KMC results seems to be reasonable.

A more detailed comparison of our simulations with experiments by
Mathon et al.  \cite{Mathon-97} is hindered by a low resolution in
measurements of precipitate sizes $R$ by SANS method used in
\cite{Mathon-97}: $\Delta R\sim 0.5$ nm. It can explain quantitative
disagreements between the results reported in \cite{Mathon-97} and
those shown in frames 2(e) and 2(f). The maximum value of the
precipitate density  $d_p(t)$ reported in \cite{Mathon-97},
$d_{max}\sim 1.65\times 10^{23}1/{\rm m}^3$, is by about 1.5 times
lower than the final values $d_p(t)$ shown in Figs. 2(e) and 2(f),
even though the maximum value $d_{max}$ is probably not reached yet
in these simulations. Hence, our simulated $d_p$ exceed those
reported in \cite{Mathon-97} by at least several times. At the same
time, values of the mean precipitate radius for the nucleation stage
(which corresponds to $d_p(t)<d_{max}$) reported in
\cite{Mathon-97}, $R_{m,M}^n\sim 1.1$---$1.65$ nm, are by about 2-3
times higher than the analogous values $R_m^n\sim 0.3-0.5$ nm
observed in both experiments \cite{Goodman-73,Kampmann-86} and
simulations \cite{SF-07,KVZ-13} for other Fe-Cu alloys, while our
$R_m$ in frames 2(e) and 2(f) are close to these usual $R_m^n$.
Therefore, it seems probable that because of the above-mentioned low
resolution $\Delta R\sim 0.5$ nm, values of the precipitate density
$d_p$ reported in \cite{Mathon-97} are underestimated, while those
of the mean precipitate size $R_m$ are overestimated. More accurate
measurements of $d_p$ and $R_m$ in dilute Fe-Cu alloys under
electron irradiation are evidently needed for a quantitative
comparison with our simulations.

\section{Estimates of effective diffusivity of copper atoms for growth
of a precipitate using kinetic Monte Carlo simulations}

In the phenomenological theory of growth of a supercritical embryo
developed for a spherical embryo of a radius $R$ with a sharp edge,
the temporal dependence  $R(t)$ (for the rigid lattice alloy model
used) is described by the following equation \cite{Lif-Pit-79}:
\begin{equation}
dR/dt=D_{eff}(R-R_c)(c-c_b)/R^2. \label{dR/dt}
\end{equation}
Here $D_{eff}$ is the effective diffusivity of a solute atom (in our
case, of a copper atom) in an alloy; $R_c$ is the critical radius;
$c=\,x$ is the copper concentration far from the embryo; and $c_b$
is the binodal concentration (solubility limit) for temperature
under consideration.

In more realistic descriptions, in particular, in the STN or KMC
simulations described in sections 2 and 3, both critical and
supercritical embryos have not a sharp edge but a diffuse surface
illustrated by Fig. 1. More important,  in the phenomenological
derivation \cite{Lif-Pit-79} of Eq. (\ref{dR/dt}), the diffusivity
$D_{eff}$ was supposed to be constant independent of the precipitate
size $N$ and its mobility $m(N)$. At the same time, the recent KMC
studies \cite{SF-07,Jourdan-10} have shown that the small
precipitates under consideration which contain $N\lesssim$100 copper
atoms are highly mobile, being much more mobile than individual
copper atoms. It leads to a great acceleration of precipitation
kinetics which qualitatively corresponds to an increase of the
effective diffusivity $D_{eff}$ in Eq. (\ref{dR/dt}) with respect to
the diffusivity of an individual copper atom.

In spite of all these simplifications, the phenomenological equation
(\ref{dR/dt}) is commonly believed to realistically describe growth
of supercritical embryos \cite{Lif-Pit-79}. Accepting this point of
view, we first present the explicit solution of Eq. (\ref{dR/dt})
for $R(t)$. Then we use our KMC simulations to access validity of
Eq. (\ref{dR/dt}) and to estimate the parameter $D_{eff}$ in this
equation for dilute Fe-Cu alloys under consideration.

First we note that employing the simplest sharp-edge model of an
embryo in usual derivations of Eq. (\ref{dR/dt}) seems to be
unessential, and this equation (at large $N_c\gg 1$ considered)
appears to be valid for any realistic description of growth of an
embryo, including that used in KMC simulations. The radius $R(t)$ in
Eq. (\ref{dR/dt}) mainly characterizes the total number $N$ of
solute atoms within the embryo, and for any precipitate this radius
can be expressed via $N$  analogously to Eq. (\ref{R_c}) for the
critical embryo:
\begin{equation}
R=a_0(3N/8\pi)^{1/3}=0.142\,N^{1/3}\,{\rm nm}. \label{R-N}
\end{equation}
The right-hand-side of Eq. (\ref{dR/dt}) contains two basic factors
(in addition to the diffusivity $D_{eff}$ and a geometrical factor
$1/R^2$) which naturally describe the driving force for growth of
the embryo: the factor $(c-c_b)$ that characterizes supersaturation
of a metastable alloy, and the factor $(R-R_c)$ that describes
vanishing of this force at $R=R_c$. Therefore, the phenomenological
equation (\ref{dR/dt}) with the generalization (\ref{R-N}) seems to
be rather plausible, and comparison to our KMC simulations given
below seems to confirm its validity, at least for the $R$ and $t$
intervals studied.

First we present the explicit solution $R(t)$ of Eq. (\ref{dR/dt}).
As discussed in detail in \cite{Lif-Pit-79}, the embryo can be
considered as ``supercritical'' only when its size $R$ exceeds some
value $(R_c+\Delta)$ where $\Delta$ determines the scale of critical
fluctuations of sizes near $R_c$. Therefore, the time $t_{f}$ when
the supercritical embryo has been eventually formed is defined by
the relation:
\begin{equation}
R(t_{f})=(R_c+\Delta).\label{t_fs}
\end{equation}
For the sharp-edge model of the embryo used in \cite{Lif-Pit-79},
the fluctuation width $\Delta$ is expressed via the interfacial
energy $\sigma$ as: $\Delta=(T/8\pi\sigma)^{1/2}$. To generalize
this estimate to the case of real embryos with diffuse interfaces,
we can express $\sigma$ via the critical radius $R_c$ and the
nucleation barrier $F_c$. It yields:
\begin{equation}
\Delta= R_c\alpha,\qquad \alpha=(2T/3F_c)^{1/2}.\label{alpha}
\end{equation}
For large embryos under consideration with $\beta F_c\gg 1$, values
of $\alpha$ are small; in particular, for the Fe-Cu alloys listed in
tables 1 and 2, we have: $\alpha\sim 0.15-0.2$.

Integrating Eq. (\ref{dR/dt}) from the initial time $t_{f}$ defined
by Eq. (\ref{t_fs}) to the arbitrary time $t$ when $R(t)=R$, we
obtain:
\begin{equation}
(R-R_c-\Delta)(R+3R_c+\Delta)/2+R_c^2\ln\,[(R-R_c)/\Delta]=
D_{eff}(c-c_b)\tau\label{R(t)}
\end{equation}
where $\tau=(t-t_{f})$ is the total  time of growth of a
supercritical embryo. If we describe this growth by a reduced
variable $y=(R-R_c-\Delta)/R_c$, Eq. (\ref{R(t)}) takes a universal
form which contains only the reduced dimensionless time $\xi$:
\begin{eqnarray}
&&y(2+\alpha+y/2)+\ln(1+y/\alpha)=\xi\nonumber,\\
&&\xi=D_{eff}(c-c_b)\tau/R_c^2.\label{y(t)}
\end{eqnarray}
At large $R\gg R_c$, the dependence $R(\tau)$ given by Eq.
(\ref{R(t)}) or (\ref{y(t)}) takes the form
\begin{equation}
R(\tau)=[2D_{eff}(c-c_b)\tau]^{1/2}\label{R_diff-tau}
\end{equation}
which describes the diffusion-controlled growth of a spherical
embryo, see, e. g., \cite{Martin-76a}.

Now we compare the description of growth of an embryo by the
phenomenological equation (\ref{R(t)}) or (\ref{y(t)}) with that
given by our KMC simulations. First, we note that for our model of
irradiated alloys based on Eqs. (\ref{D_Cu}) and (\ref{t_irr}), the
reduced time $\xi$ in Eq. (\ref{y(t)}) does not include the
irradiation acceleration factor $A_{irr}$,\, just as the right-hand
side of Eq. (\ref{F_c-KMC}), and this reduced time is explicitly
expressed via the thermal aging time $\tau=\tau_{th}=
(t_{th}-t_{th}^{f})$ used in our KMC simulations. Therefore, growth
of a supercritical embryo can be described, from one side, by the
function $R(\xi)$ determined by Eq. (\ref{R(t)}) or (\ref{y(t)}).
From the other side, such growth can be followed in our KMC
simulations for any precipitate chosen, which yields the dependence
$R=R_{\rm KMC}(t_{th})$ with $R$ defined by Eq. (\ref{R-N}).
Equating these two quantities,
\begin{equation}
R(\xi)=R_{\rm KMC}(t_{th}),\label{R_xi_KMC}
\end{equation}
with $\xi$ given by Eq. (\ref{y(t)}) for $\tau=\tau_{th}$, we obtain
the equality which contains only one unknown parameter $D_{eff}$ at
all times $\tau_{th}>0$.

One can expect that the proper choice of this single parameter can
provide a good accuracy for obeying equality (\ref{R_xi_KMC}) at all
$\tau_{th}$  considered only if the phenomenological equation
(\ref{dR/dt}) holds true for these $\tau_{th}$. Therefore, the check
of validity of relation (\ref{R_xi_KMC}) enables us, first, to
access reliability of the phenomenological equations (\ref{dR/dt})
and (\ref{R(t)}) for description of growth of supercritical embryos
and, second, to estimate the effective diffusivity $D_{eff}$ in
these equations  for dilute Fe-Cu alloys.

It is convenient to characterize the effective diffusivity
$D_{eff}(x,T)$ by its ratio $A_c$ to the appropriate dilute alloy
diffusivity $D(0,T)=D_0(T)$ writing $D_{eff}$ as
\begin{equation}
D_{eff}(x,T)=A_c\times D_0(T).\label{A_c}
\end{equation}
The parameter $A_c=A_c(x,T)$ in (\ref{A_c}) characterizes the
effective acceleration of growth of precipitates being mainly due to
the above-mentioned high mobility of small copper clusters. To
differ it from the irradiation acceleration factor $A_{irr}$
discussed in Sec. 4, $A_c$ will be called ``the acceleration
diffusion parameter''.

For the dilute alloy diffusivity $D_0(T)$  of a copper atom in a
pure iron at $T=290-450^{\circ}$C we will use the conventional
Arrhenius-type expression
\begin{equation}
D_0(T)=A\exp\,(-\beta Q)\label{D_0}
\end{equation}
with the $A$ and $Q$ values suggested by Soisson and Fu \cite{SF-07}
basing on combinations of their $ab \ initio$ calculations and
empirical estimates:
\begin{equation}
A=0.97\cdot 10^{-4} \ {\rm m^2/s}\,,\qquad Q=2.67 \ {\rm eV}.
\label{SF-parameters}
\end{equation}

In Figs. 5 and 6 we show the phenomenological functions $R(\xi)$
defined by Eqs. (\ref{R(t)}) and (\ref{y(t)}) together with the
dependencies $R_{\rm KMC}(t_{th})$ observed in our KMC simulations.
Each curve  $R_{\rm KMC}(t_{th})$ in these figures starts from the
time of formation of the precipitate chosen observed in the KMC
simulation, while the curve $R(\xi)=R[\xi(t_{th})]$ starts from the
time $t_{th}^{f}$ of formation of the supercritical embryo defined
by Eq. (\ref{t_fs}). Each frame in these figures corresponds to some
of our simulations; for other simulations (not shown in Figs. 5 and
6), the results are similar. The parameter $A_c$ and the time
$t_{th}^{f}$ used to draw the curve $R(\xi)$ in each frame have been
estimated from the best fit of this $R(\xi)$ to the $R_{\rm
KMC}(t_{th})$ curve at this frame. In Fig. 7 we show the
concentration and temperature dependencies of the acceleration
diffusion parameter $A_c$ obtained in these KMC estimates.

Let us discuss the results presented in Figs. 5\,-\,7. First, we
note that some of irregularities in dependencies $R_{\rm
KMC}(t_{th})$ seen in Figs. 5 and 6 can be related not to the real
statistical fluctuations of sizes but to the methodical errors due
to too long temporal intervals between sequent savings of KMC data.
This seems to be the case, in particular, for initial stages of
simulations shown in frames 5(a), 6(c) and 6(d), and also for some
other  frames, e. g., 6(a). At the same time, for the later stages
of growth when the precipitate becomes supercritical, $R\gtrsim
(R_c+\Delta)$, these methodical distortions seem to be less
significant. Second, Figs. 5 and 6 clearly illustrate the strong
fluctuations of sizes in the ``critical'' region $R\lesssim
(R_c+\Delta)$ mentioned in the discussion of Eqs. (\ref{t_fs}) and
(\ref{alpha}). Such fluctuations are pronounced, in particular, in
frames  5(b), 5(c), 6(c), 6(e), and 6(f). Third and most important,
Figs. 5 and 6 show that the description of growth of precipitates by
the phenomenological equations (\ref{dR/dt}) and (\ref{R(t)}) seems
to agree well with the KMC simulations, at least up to $R\gtrsim 1.5
R_c$, that is, under increase of the number of copper atoms in the
growing embryo  by about four times.

Figs. 5 and 6 also clearly illustrate the effects of cluster
mobility and their direct coagulation mentioned above. In
particular, sudden increases (``jumps'') of $R(t)$ seen in Fig. 5(d)
at $t_{th}=10.6\times 10^{9}$ s, as well as in Fig. 5(f) at
$t_{th}=1.5\times 10^{9}$  and $t_{th}=1.75\times 10^{9}$ s, occur
simultaneously with the disappearance (evidently, due to
coagulation) of one of supercritical clusters in Figs. 2(b) and
2(f). Similarly, jumps of $R(t)$ in Fig. 6(a) at $t_{th}=6.56\times
10^{9}$ s and in Fig. 6(b) at $t_{th}=1.46\times 10^{9}$ s  occur
simultaneously with the disappearance of one supercritical cluster
in Figs. 3(b) and 3(d), respectively. The smaller jumps in $R(t)$
seen in Figs. 5(d), 5(e), 6(a) and 6(b) can correspond to
coagulation of subcritical clusters which are not registered in
Figs. 2 and 3. At the same time,  for simulations with $N_p=1$ shown
in Figs. 5(a)-5(c) and 6(d)-6(f), for which only one supercritical
cluster is present in the simulation box, the large and distinct
jumps in $R(t)$ are not clearly seen. Inspection of atomic
distributions for these simulations shows that growth of this
supercritical cluster is mainly realized via its fast diffusion
among almost immobile individual copper atoms which are sometimes
``swept'' and absorbed by this mobile cluster. Some subcritical
clusters containing several copper atoms are also observed in these
simulations, and they seem to diffuse notably slower than the
supercritical cluster, in a qualitative agreement with Fig. 9 in
\cite{SF-07}.

The resulting acceleration diffusion parameter $A_c$ for dilute
Fe-Cu alloys is shown in Fig. 7. It is rather high: $A_c\gtrsim
200$, having the same scale as the parameters of acceleration of
precipitation in Fe-1.34\%Cu alloys at $T=500^{\circ}$C due to the
high mobility of copper clusters studied  in Refs. \cite{SF-07} and
\cite{Jourdan-10}. Fig. 7 also shows that the concentration and
temperature dependencies of this acceleration are rather sharp. Fig.
7(a) illustrates the notable increase of $A_c$ with increase of
copper concentration $x$; it can be explained by a probable
enhancement of density of mobile copper clusters (both supercritical
and subcritical) under increase of the copper content in an alloy.
Fig. 7(b) shows a significant decrease of $A_c$ under elevating
temperature $T$. This also seems natural, as elevating $T$ should
lead to the weakening of the copper-vacancy binding proportional to
the Mayer functions $[\exp\,(\beta E^{bn}_{v{\rm Cu}})-1]$ with
$E^{bn}_{v{\rm Cu}}$ from (\ref{E^b-SF}) \cite{KSSV-11}. Hence the
strong trapping of vacancies by copper clusters (which is the
physical origin of their high mobility \cite{SF-07}) should weaken.

\section{Comparison of nucleation barriers found
using the STN calculations and the KMC simulations}

Estimates of the nucleation rate $J_{\rm KMC}$ described in Sec. 3
and illustrated by Figs. 2 and 3, combined with the estimates of the
effective diffusivity $D_{eff}$ discussed in Sec. 5 and illustrated
by Figs. 5-7, enable us to find the reduced nucleation barriers
$\beta F_c$ in Eq. (\ref{F_c-KMC}) basing on the kinetic Monte Carlo
simulations. This equation (\ref{F_c-KMC}) also shows that for the
large $\beta F_c$ considered which vary with $x$ and $T$ very
sharply, possible errors of a relative order of unity in our
estimates of parameters $\tilde{J}_0$, $J_{\rm KMC}$ and $D_{eff}$
make no significant effect on the $\beta F_c$ values obtained.

In table 4 we present the $(\beta F_c)_{\rm KMC}$ values estimated
for all simulations shown in Figs. 2 and 3, together with the
analogous $(\beta F_c)_{\rm STN}$ values  calculated in Sec. 2 with
the use of the statistical theory of nucleation
\cite{DV-98a,DV-98b}. In comparison of these KMC and STN results we
should remember that the STN assumes the size and the nucleation
barrier of the embryo to be large: $N_c\gg 1,\ \beta F_c\gg 1$.
Therefore, the accuracy of this theory and its agreement with the
KMC simulations should improve when the size of the embryo
increases. The results presented in tables 1, 2 and 4 agree with
these considerations. We see that at low concentrations $x\leq
0.2\%$ when both $N_c$ and $\beta F_c$ are sufficiently large:
$N_c\gtrsim 12$, $\beta F_c\gtrsim 20$, the agreement between KMC
and STN results is virtually perfect: for all nine simulations with
$x\leq 0.2\%$ shown in table 4, differences between the $(\beta
F_c)_{\rm KMC}$ and the $(\beta F_c)_{\rm STN}$ values have the
order of a percent. At the same time, at higher $x$ equal to 0.3 or
0.4\%, the $N_c$ and $\beta F_c$ values notably decrease, and
differences between $(\beta F_c)_{\rm KMC}$ and $(\beta F_c)_{\rm
STN}$ increase, which may reflect the lowering of accuracy of the
STN.

Therefore, the results presented in table 4 enable us to make an
important conclusion that  at high values $\beta F_c\gtrsim 20$, the
STN-based calculations of nucleation barriers are highly reliable.
At the same time, as mentioned in Sec. 2, the homogeneous
precipitation limit in dilute Fe-Cu alloys corresponds just to the
values $\beta F_c> 20$. Therefore, the nucleation barriers near the
homogeneous precipitation limits in these alloys can be reliably
calculated using the STN.

\section{Temperature dependencies of precipitation limits
in dilute iron-copper alloys}

Calculations of nucleation barriers described in Secs. 2 and 6 can
be used for tentative estimates of the homogeneous precipitation
limit $x_l^{hom}(T)$ at different temperatures $T$. As mentioned in
Sec. 2, it seems natural to suggest that the position of this limit
is mainly determined by the value of the activation factor $\exp
(-\beta F_c)$, that is, by the reduced nucleation barrier $\beta
F_c$ which at $x=x_l^{hom}$ takes a certain high value $(\beta
F_c)_l$. It seems also natural to assume that these precipitation
limits at different temperatures $T$ correspond to the similar
values of the activation factor $\exp (-\beta F_c)$. It implies the
following relation:
\begin{equation}
(\beta F_c)_l\equiv\beta F_c\Big[x_l^{hom}(T),T\Big]\simeq
C_{hom}\label{F_cl-hom}
\end{equation}
where $C_{hom}$ is a constant independent of temperature, and thus
it can be estimated using the experimental value\,
$x_l^{hom}(290^{\circ}{\rm C})\equiv x_l^{exp}$.\, For example,
using  estimates  of $x_l^{exp}$ mentioned in Sec. 2,\,
$x_l^{exp}\sim 0.08-0.09\%$ from \cite{Mathon-97}  or $x_l^{exp}\sim
0.15\%$  from \cite{Auger-00}, we obtain:  $C_{hom}\simeq 28$ or
$C_{hom}\simeq 23 $. Then equation (\ref{F_cl-hom}) with the
function $F_c(x,T)$ calculated using the STN enable us to find the
$x_l^{hom}(T)$ value. A high accuracy of the STN calculations of
nucleation barriers combined with the above-mentioned physical
considerations allow us to expect that such ``semi-empirical''
estimates of the homogeneous precipitation limit for temperatures
$T$ of practical interest can be sufficiently reliable.

One can also try to extend this approach to the case of the
heterogeneous precipitation of copper on clusters of
irradiation-induced point defects, vacancies and self-interstitial
atoms (that is, to the case of neutron or ion irradiation
\cite{Akamatsu-95}-\cite{Meslin-10b}), to estimate the appropriate
heterogeneous precipitation limit $x_l^{het}$, which is more
interesting for applications. Even though kinetic paths of the
heterogeneous and the homogeneous precipitation differ from each
other (see, e. g., \cite{Martin-76b}), the main physical origin for
suppressing precipitation at low solute concentrations $x$ seems to
be the same  for both processes. In both cases, formation of a
precipitate needs overcoming the energetic barrier due to the
surface loss in the free energy $F$ which is not compensated by the
volume gain in  $F$ until the precipitate volume becomes
sufficiently large. Differences between these two processes are
usually related mainly to some geometrical factors which depend on
the structure of the heterogeneity \cite{Martin-76b}.

Therefore, one may expect that the heterogeneous precipitation limit
$x_l^{het}$  can also be estimated from a phenomenological relation
similar to (\ref{F_cl-hom}), but the ``limiting'' value $F_{cl}$ in
this relation should be  higher as the nucleation barrier for the
heterogeneous nucleation, $F_c^{het}$, is lower than that for the
homogeneous nucleation, $F_c^{hom}$. However, if we assume that the
difference between $F_c^{het}$ and $F_c^{hom}$ has mainly a
geometrical origin (which is the case for the simplest models of
heterogeneous nucleation \cite{Martin-76b}), then the ratio
$F_c^{het}/F_c^{hom}$ can be supposed to weakly vary with
temperature. Then the temperature dependence of the heterogeneous
precipitation limit $x_l^{het}(T)$ can be estimated from a
semi-empirical relation similar  to (\ref{F_cl-hom}):
\begin{equation}
\beta F_c\Big[x_l^{het}(T),T\Big]\simeq C_{het}\label{F_cl-het}
\end{equation}
where $F_c(x,T)$ is again the STN calculated nucleation barrier for
the homogeneous nucleation, while  the constant $C_{het}$ is
estimated using the experimental value\, $x_l^{het}(290^{\circ}{\rm
C})\equiv x_{l,exp}^{het}$.\, For numerical estimates we use the
value \,$x_{l,exp}^{het}\simeq 0.06\%$ given in Refs.
\cite{Miller-06} and \cite{Miller-09}, which yields:\, $
C_{het}\simeq 35$.

In figure 8 we show positions of the homogeneous and the
heterogenous precipitation limits in the $(x,T)$ plane,
$T_l^{hom}(x)$ and $T_l^{het}(x)$, calculated for dilute Fe-Cu
alloys using Eqs. (\ref{F_cl-hom}) and (\ref{F_cl-het}). For
comparison, in Fig. 8 we also show the binodal curve $T_b(x)$ (the
copper solubility limit) estimated in \cite{SF-07} from experimental
data and $ab \ initio$ calculations. The results in Fig. 8 are
presented for the temperature interval between 290 and
450$^{\circ}$C used for annealing of RPVs \cite{Auger-00,Miller-09}.
The value $C_{hom}=27.6$ used in Fig. 8 corresponds to
$x_l^{hom}(290^{\circ}{\rm C})=0.088\%$ for which Radiguet et al.
\cite{Radiguet-07} did not observe any homogeneous precipitation,
while the constant $C_{het}=35$ used in Fig. 8 corresponds to the
estimate \,$x_l^{het}= 0.06\%$ given by Miller et al.
\cite{Miller-06,Miller-09}. Our calculations also show that a slight
variation of the constant $C_{hom}$  or $C_{het}$ leads to an
``almost rigid'' shift of the curve $T_l^{hom}(x)$ or $T_l^{het}(x)$
in Fig. 7. For example, when this variation of $C_{hom}$ leads to
the shift $\delta x_l^{hom}$=0.05\% to the right at $290^{\circ}$C,
the analogous shift at $450^{\circ}$C is almost the same: $\delta
x_l^{hom}(450^{\circ}{\rm C})\simeq 0.06\%$.

The most interesting general feature seen in Fig. 8 seems to be a
rather sharp temperature dependence of the precipitation limits,
particularly for the heterogeneous precipitation. For example,  the
$x_l^{het}$ value at $T=450^{\circ}$C exceeds that at
$T=290^{\circ}$C by almost five times, while the $x_l^{hom}$  value
increases for this temperature interval by about 3.5 times. These
qualitative conclusions can be useful, in particular, for the
interpretation of microstructural observations related to the
annealing of RPVs \cite{Miller-09}. Note also that the variations of
both precipitation limits, $x_l^{het}(T)$ and $x_l^{hom}(T)$, with
temperature for the temperature interval considered are much
stronger than those for the solubility limit $x_b(T)$.

To conclude this section, we again note that our equations
(\ref{F_cl-hom}) and (\ref{F_cl-het}) for temperature dependencies
$x_l^{het}(T)$ and  $x_l^{hom}(T)$ are of course tentative and have
not been proved formally. At the same time, physical considerations
about the dominant role of the activation factor $\exp\,(-\beta
F_c)$ in suppressing precipitation at low $x$ used in our derivation
seem to be rather plausible, particularly for the homogeneous
precipitation. In the  derivation of Eq. (\ref{F_cl-het}) for the
heterogeneous precipitation,  we also used an assumption about a
weak temperature dependence of the ratio $F_c^{het}/F_c^{hom}$ for
the temperature interval considered which may seem to be less
evident. Therefore, the accuracy of Eq. (\ref{F_cl-het}) for
$x_l^{het}(T)$ can, generally, be lower than the accuracy of Eq.
(\ref{F_cl-hom}) for $x_l^{hom}(T)$. However, one may expect that
the main qualitative features of both dependencies, $T_l^{hom}(x)$
and $T_l^{het}(x)$, shown in Fig. 8 are correctly described by the
simple model used.

\section{Conclusions}

Let us summarize the main results of this work. We study  kinetics
of homogeneous nucleation and growth of copper precipitates under
electron irradiation of iron-copper alloys at low concentrations
$x=0.06-0.4$ at.\% and temperatures $T=290-450^{\circ}$C used for
service of a number of nuclear reactor pressure vessels
\cite{Akamatsu-95}-\cite{Meslin-10b}. The earlier-described kinetic
Monte Carlo (KMC) modeling \cite{SF-07} and the statistical theory
of nucleation (STN) \cite{DV-98a,DV-98b}  are used. The $ab \
initio$ model of interatomic interactions which describes well the
available data about precipitation in Fe-Cu alloys at different $x$
and $T$ \cite{SF-07,KVZ-13} is used for both the KMC simulations and
the STN calculations. The conventional assumption \cite{Mathon-97}
about the similarity of mechanisms of precipitation under electron
irradiation and under thermal aging is also adopted. Then
precipitation under electron irradiation can be described by the KMC
codes developed by Soisson and Fu \cite{SF-07} for studies of
precipitation during thermal aging but with the acceleration of
kinetics under irradiation  characterized by the acceleration factor
$A_{irr}$ in Eqs. (\ref{D_Cu}) and (\ref{t_irr}). We estimate this
acceleration factor for dilute iron-copper alloys considered, and
our estimate reasonably agrees with the available experimental data
\cite{Mathon-97}.

Our STN-based calculations of  the nucleation barrier $F_c$  in the
Zeldovich-Volmer relation (\ref{J-J_0}) for the nucleation rate $J$
show that this nucleation barrier varies with $x$ and $T$ very
sharply. Thus the concentration and temperature dependencies of the
nucleation rate are mainly determined by the variations of the
activation factor $\exp\,(-\beta F_c)$ with $x$ or $T$. We also
found that at $T=290^{\circ}$C, the interval of concentrations $x$
for which this activation factor starts to fall off very rapidly
just corresponds to the interval of positions of the homogeneous
precipitation limit $x_l^{hom}$ estimated in experiments
\cite{Mathon-97,Auger-00,Radiguet-07}.

Our KMC simulations for the dilute alloys considered describe mainly
the very initial stage of precipitation, that of the steady-state
nucleation characterized by the constant nucleation rate $J$. We
also use these KMC simulations to study kinetics of growth of a
supercritical embryo and confirm the validity of the
phenomenological equation of the classical theory of nucleation
which describes this growth \cite{Lif-Pit-79}.  Our simulations also
enable us to estimate the effective diffusivity $D_{eff}$ which
enters into the STN expression for the nucleation rate $J$. The
$D_{eff}$ values are found to exceed the dilute alloy values
$D_0(T)$ by two-three orders of magnitude. This effective
acceleration of diffusion seems to be mainly  due to the high
mobility of small copper clusters found earlier for the
precipitation in Fe-1.34\%Cu alloy at $T=500^{\circ}$C
\cite{SF-07,Jourdan-10}.

The KMC estimates of the nucleation rate $J$ and the effective
diffusivity $D_{eff}$ described above enable us  to determine the
nucleation barrier $F_c$ using Eq. (\ref{F_c-KMC}) based on the KMC
simulations. The resulting values of the reduced nucleation barrier
$(\beta F_c)_{\rm KMC}$ are compared with the $(\beta F_c)_{\rm
STN}$ values calculated using the statistical theory of nucleation
\cite{DV-98a,DV-98b}.  We find that for the dilute alloys with
$x\leq 0.2\%$, the STN and the KMC results for the nucleation
barrier $F_c$ coincide within about a percent. It seems to confirm a
high reliability of the STN for this problem.

Making a plausible assumption that the position of the homogeneous
nucleation limit $x_l^{hom}$ is mainly determined by the value of
the activation factor $\exp\,[-\beta F_c(x,T)]$ which at
$x=x_l^{hom}$ takes an approximately same value $\exp\,(-\beta
F_c)_l$ for all temperatures $T$ considered, we use our STN-based
calculations of reduced nucleation barriers $\beta F_c$ to estimate
the temperature dependence $x_l^{hom}(T)$. For temperatures between
290 and 450$^{\circ}$C, this dependence is found to be rather sharp,
much sharper than that for the solubility limit $x_{sol}(T)$. Making
an additional assumption about a weak temperature dependence of the
ratio of nucleation barriers for the heterogeneous and the
homogeneous nucleation, $F_c^{het}/F_c^{hom}$, we also estimate the
temperature dependence of the heterogeneous precipitation limit
$x_l^{het}(T)$ which corresponds to the neutron or ion irradiations.
The dependence $x_l^{het}(T)$ is found to be still more sharp than
that for the homogeneous precipitation. In spite of the evidently
tentative character of these estimates, one can expect that the main
qualitative features of both dependencies, $x_l^{hom}(T)$ and
$x_l^{het}(T)$, are described by this model properly.

\section*{Acknowledgements}

We are very grateful to V. Yu.  Dobretsov for his help in the
computations using the STN, as well as to A. Barbu and K. Yu.
Khromov, for fruitful discussions. The work was supported by the
Russian Fund of Basic Research (grant No. 12-02-00093); by the fund
for support of leading scientific schools of Russia  (grant No.
NS-215.2012.2); and by the program of Russian university scientific
potential development (grant  No. 2.1.1/4540).

\newpage

\vskip5mm \vbox{\noindent Table 1. Some characteristics of critical
embryos in alloys Fe$_{1-x}$Cu$_x$ at $T=290^{\circ}$C calculated
using the ``pair-cluster-with-fluctuations''  approximation of STN
\cite{DV-98a} and Cu-Cu interactions given by Eqs. (\ref{E^b-SF}).
$N_c$ is the number of copper atoms within the embryo, and $R_c$  is
its effective radius defined by Eq. (\ref{R_c}); $\Delta\Omega_0$
and $\Delta\Omega_1$ are the nucleation barrier values found in the
zeroth and the first order in fluctuations, as explained in the
text, and $F_c=\Delta\Omega_0 +\Delta\Omega_1$ is the full
nucleation barrier.
\vskip5mm
\begin{tabular}{|c|ccccccccccc|}
\hline
&&&&&&&&&&&\\
$x$, {\%} &0.06&0.07&0.08&0.088&0.1&0.116&0.144&0.172&0.2&0.3&0.4\\
\hline
&&&&&&&&&&&\\
$N_c$&25.2&23.1&20.6&19.3&17.8&16.3&14.5&13.2&12.1&9.5&7.9\\
&&&&&&&&&&&\\
$R_c$, nm&0.42&0.40&0.39&0.38&0.37&0.36&0.35&0.34&0.33&0.30&0.28\\
\hline
&&&&&&&&&&&\\
$\beta\Delta\Omega_0$&28.9&25.3&22.6&20.8&18.6&16.3&13.2&11.1&9.4&5.8&3.6\\
&&&&&&&&&&&\\
$\beta\Delta\Omega_1$&5.9&5.6&6.1&6.8&7.7&8.7&9.8&10.3&10.4&9.6&8.4\\
\hline
&&&&&&&&&&&\\
$\beta F_c$&34.9&30.9&28.7&27.6&26.3&25.0&23.0&21.4&19.8&15.4&12.2\\
\hline

\end{tabular}}

\vskip10mm \vbox{\noindent Table 2. The same as in table 1 but  at
$x$=0.2\% and different temperatures $T$.
 \vskip5mm
\begin{tabular}{|c|cccccccc|}
\hline
&&&&&&&&\\
$T$, $^{\circ}$C  &290&305&320&335&350&365&380&390\\
\hline
&&&&&&&&\\
$N_c$&12.1&13.4&14.8&16.5&18.4&20.8&24.1&27.2\\
&&&&&&&&\\
$R_c$(nm)&0.33&0.34&0.35&0.36&0.37&0.39&0.41&0.43\\
\hline
&&&&&&&&\\
$\beta\Delta\Omega_0$&9.4&10.9&12.5&14.2&16.1&18.2&20.7&22.5\\
&&&&&&&&\\
$\beta\Delta\Omega_1$&10.4&9.9&9.1&8.2&7.2&6.2&5.5&5.5\\
\hline
&&&&&&&&\\
$\beta F_c$&19.8&20.8&21.6&22.4&23.3&24.5&26.2&28.0\\
\hline

\end{tabular}}

\vskip10mm \vbox{\noindent Table 3. Values  of various factors in
the expression (\ref{J_0}) for the prefactor $J_0$ in Eq.
(\ref{J-J_0}) for some Fe$_{1-x}$Cu$_x$  alloys; $\gamma_{nn}$ is
the effective exchange rate between neighboring copper and iron
atoms given by  Eq. (\ref{gamma_nn}), and $v_a=a_0^3/2$ is volume
per atom in the BCC lattice of iron.
\vskip5mm
\begin{tabular}{|c|ccccc||c|ccc|}
\hline
&&&&&&&&&\\
$T$&\multicolumn{5}{c||}{$290^{\circ}$C}&$x$&\multicolumn{3}{c|}{0.2 \%}\\
\hline
&&&&&&&&&\\
$x$, {\%} &0.06&0.1&0.2&0.3&0.4&$T$, $^{\circ}$C&290&335&390\\
\hline
&&&&&&&&&\\
$\beta\gamma_0$&-2.0&-3.9&-5.0&-5.5&-6.0&&-5.0&-3.9&-1.8\\
&&&&&&&&&\\
${\cal N}$&1.3&1.6&1.7&2.2&2.8&&1.7&1.4&1.1\\
&&&&&&&&&\\
$D_{aa}/\gamma_{nn}$&0.12&0.13&0.14&0.13&0.11&&0.14&0.15&0.15\\
&&&&&&&&&\\
$v_a D_{\bf R}({\bf u})$&5.6&3.7&1.1&0.46&0.22&&1.1&2.7&6.1\\
\hline
&&&&&&&&&\\
$\tilde{J_0}=v_aJ_0/\gamma_{nn}$&0.54&0.60&0.23&0.12&0.07&&0.23&0.45&0.57\\
\hline
\end{tabular}}

\vskip10mm \vbox{\noindent Table 4. Values  of the reduced
nucleation barrier in Fe-Cu alloys estimated in our KMC simulations,
$(\beta F_c)_{\rm KMC}$, and those calculated using the STN, $(\beta
F_c)_{\rm STN}$, at different temperatures $T$, concentrations $x$,
and simulation sizes $L_n$ in (\ref{L_n}).
\vskip5mm
\begin{tabular}{|c|cc|cc|cc|c|}
\hline
&\multicolumn{7}{c|}{}\\
$T$&\multicolumn{7}{c|}{$290^{\circ}$C}\\
\hline
&&&&&&&\\
$x$, {\%}
&\multicolumn{2}{c|}{0.172}&\multicolumn{2}{c|}{0.2}&\multicolumn{2}{c|}{0.3}&0.4\\
\hline
&&&&&&&\\
$L$&$L_2$&$L_3$&$L_1$&$L_3$&$L_1$&$L_3$&$L_1$\\
\hline
&&&&&&&\\
$(\beta F_c)_{\rm KMC}$&21.4&21.9&20.1&20.0&17.9&17.9&16.8\\
\hline
&&&&&&&\\
$(\beta F_c)_{\rm STN}$&21.4&21.4&19.8&19.8&15.4&15.4&12.2\\
\hline
\end{tabular}}
\begin{tabular}{|c|cc|ccc|cc|}
\hline
&\multicolumn{7}{c|}{}\\
$x$, \%&\multicolumn{7}{c|}{0.2}\\
\hline
&&&&&&&\\
$T$, $^{\circ}$C
&\multicolumn{2}{c|}{290}&\multicolumn{3}{c|}{305}&\multicolumn{2}{c|}{320}\\
\hline
&&&&&&&\\
$L$&$L_1$&$L_3$&$L_1$&$L_2$&$L_3$&$L_1$&$L_2$\\
\hline
&&&&&&&\\
$(\beta F_c)_{\rm KMC}$&20.1&20.0&21.0&22.1&21.5&22.1&21.9\\
\hline
&&&&&&&\\
$(\beta F_c)_{\rm STN}$&19.8&19.8&20.8&20.8&20.8&21.6&21.6\\
\hline
\end{tabular}

\newpage

%Figure captions to paper by Vaks et al. ``Studies of homogeneous
%precipitation...''

%
\begin{figure}
%\begin{center}
%\psfig{file=dv98_2.eps,scale=0.38, angle=0}\hspace{5mm}
%\end{center}
\caption{(color online) Concentration profile \,$\Delta
c(r_i)=c(r_i)-c$\, in the critical embryo;  $c(r_i)$=$c_i$ is the
mean occupation of site $i$ by a copper atom (local concentration)
at the distance $r_i$ from the center of embryo, and $c=x$ is the
copper concentration far from the embryo. Solid or dashed lines are
drawn to guide the eye. Curves $A$, $B$, $C$ and $D$ (red, green,
blue and grey online) correspond to the following temperatures $T$
(in $^{\circ}$C) and concentrations $x$ (in at.\%). \ $A$: $T=290$,
$x=0.06$; \ $B$: $T=290$, $x=0.2$; \ $C$: $T=290$, $x=0.3$; and $D$:
$T=390$, $x=0.2$. \label{profiles}}
\end{figure}

\begin{figure}
%\begin{center}
%\psfig{file=dv98_2.eps,scale=0.38, angle=0}\hspace{5mm}
%\end{center}
\caption{(color online) Temporal evolution of the precipitate
density $d_p=N_p/V_s$ observed in our KMC simulations with the final
number of precipitates $N_p\geq 3$. Simulation time on abscissa axis
is given in the ``thermal aging'' values $t_{th}$ related to the
observed time $t_{irr}$ by Eq. (\ref{t_irr}). Each of frames
(a)\,--\,(f) shows simulation for the following concentration $x$
(in \%), temperature $T$ (in $^{\circ}$C) and the simulation size
$L$ in $L_n$ given by Eq. (\ref{L_n}): \hbox{(a) $x$=0.172,}
$T$=290, $L=L_2$; (b) $x$=0.172, $T$=290, $L=L_3$; (c) $x$=0.2,
$T$=290, $L=L_1$; (d) $x$=0.2, $T$=290, $L=L_3$; (e) $x$=0.3,
$T$=290, $L=L_1$; and (f) $x$=0.3, $T$=290, $L=L_3$. For simulations
shown in frames (e) and (f) we also present the mean precipitate
radius $R_m$ obtained by averaging of the left-hand side of Eq.
(\ref{R-N}) over all precipitates within simulation box. Dashed
lines (blue online) show our estimates of the coefficient $J=J_{\rm
KMC}$ in Eq. (\ref{J_KMC}).\label{d_p-t-1}}
\end{figure}

\begin{figure}
%\begin{center}
%\psfig{file=dv98_2.eps,scale=0.38, angle=0}\hspace{5mm}
%\end{center}
\caption{(color online) The same as in Fig. 2 but for the following
simulations: (a) $x$=0.4, $T$=290, $L=L_1$; (b) $x$=0.2, $T$=305,
$L=L_1$;  (c) $x$=0.2, $T$=305, $L=L_2$; (d) $x$=0.2, $T$=305,
$L=L_3$; (e) $x$=0.2, $T$=320, $L=L_1$; and (f) $x$=0.2, $T$=320,
$L=L_3$.\label{d_p-t-2}}
\end{figure}

\begin{figure}
\caption{Evolution of point defect atomic fractions in pure iron,
during an irradiation at $2 \times 10^{-9}\; \rm dpa.s^{-1}$ and
$T=290^{\circ}$C, with a dislocation densities (a) $\rho _d=10^8\;
\rm cm^{-2}$ and (b) $\rho _d=10^{11}\; \rm cm^{-2}$. Solid line
(red online): vacancies; dashed line (green online): interstitial
atoms.}
\end{figure}

\begin{figure}
%\begin{center}
%\psfig{file=dv98_2.eps,scale=0.38, angle=0}\hspace{5mm}
%\end{center}
\caption{(color online) Solid lines (red online) show the
dependencies $R=R_{\rm KMC}(t_{th})$ with $R$ defined by Eq.
(\ref{R-N}) observed  in our KMC simulations for temperature
$T=290^{\circ}$C. For simulations with $N_p=1$ (those for
$x<0.172\%$), this dependence corresponds to the single precipitate
within simulation box; and for simulations with $N_p>1$ ($x\geq
0.172\%$), to the biggest precipitate within simulation box. Dashed
line (blue online) shows the dependence $R(\xi)$ given by Eq.
(\ref{y(t)}). Each of frames (a)\,--\,(f) corresponds to the
following values of the concentration $x$ (in \%), the simulation
size $L$ (in $L_n$), and the acceleration  diffusion parameter $A_c$
in Eq. (\ref{A_c}): \hbox{(a) $x$=0.088,} $L=L_1$, $A_c=334$; (b)
$x$=0.116, $L=L_1$, $A_c=396$; (c) $x$=0.144, $L=L_1$, $A_c=450$;
(d) $x$=0.172, $L=L_3$, $A_c=570$; (e) $x$=0.2, $L=L_1$, $A_c=463$;
and (f) $x$=0.3, $L=L_3$, $A_c=666$.\label{R_xi-R_KMC-1}}
\end{figure}

\begin{figure}
%\begin{center}
%\psfig{file=dv98_2.eps,scale=0.38, angle=0}\hspace{5mm}
%\end{center}
\caption{(color online) The same as in Fig. 5 but for the
concentration $x=0.2$\%. Simulations for $T\leq 320^{\circ}$C
correspond to $N_p>1$, and those for  $T> 320^{\circ}$C, to $N_p=1$.
Each of frames (a)\,--\,(f) corresponds to the following values of
temperature $T$ (in $^{\circ}$C), the simulation size $L$ (in
$L_n$), and the acceleration diffusion parameter $A_c$: \hbox{(a)
$T$=305,} $L=L_1$, $A_c=430$; (b) $T$=305, $L=L_3$, $A=350$; (c)
$T$=320, $L=L_1$, $A_c=330$; (d) $T$=335, $L=L_1$, $A_c=200$; (e)
$T$=380, $L=L_1$, $A_c=230$; and (f) $T$=390, $L=L_1$, $A_c=200
$.\label{R_xi-R_KMC-2}}
\end{figure}

\begin{figure}
%\begin{center}
%\psfig{file=dv98_2.eps,scale=0.38, angle=0}\hspace{5mm}
%\end{center}
\caption{(color online) (a) Values of the acceleration diffusion
parameter $A_c$ in Eq. (\ref{A_c}) observed in our KMC simulations
for Fe$_{1-x}$Cu$_x$ alloys at temperature  $T=290^{\circ}$C and
different concentrations $x$. Circles, squares and triangles (green,
blue and red online) correspond to the simulation size $L$ equal to
$L_1$, $L_2$ and $L_3$, respectively. (b) The same as in (a) but at
$x$=0.2\% and different temperatures $T$.\label{A_c-x,T}}
\end{figure}

\begin{figure}
%\begin{center}
%\psfig{file=dv98_2.eps,scale=0.38, angle=0}\hspace{5mm}
%\end{center}
\caption{(color online) Solid line (red online) shows the
homogeneous precipitation limit for dilute Fe-Cu alloys,
$T_l^{hom}(x)$ or $x_l^{hom}(T)$, estimated using Eq.
(\ref{F_cl-hom}) with $C_{hom}=27.6$. Dashed line (green online)
shows the heterogeneous precipitation limit for these alloys,
$T_l^{het}(x)$ or $x_l^{het}(T)$, estimated using Eq.
(\ref{F_cl-het}) with $C_{het}=35$. Chain line (blue online) shows
the binodal $T_b(x)$ as estimated in \cite{SF-07}.
\label{T_l-curves}}
\end{figure}

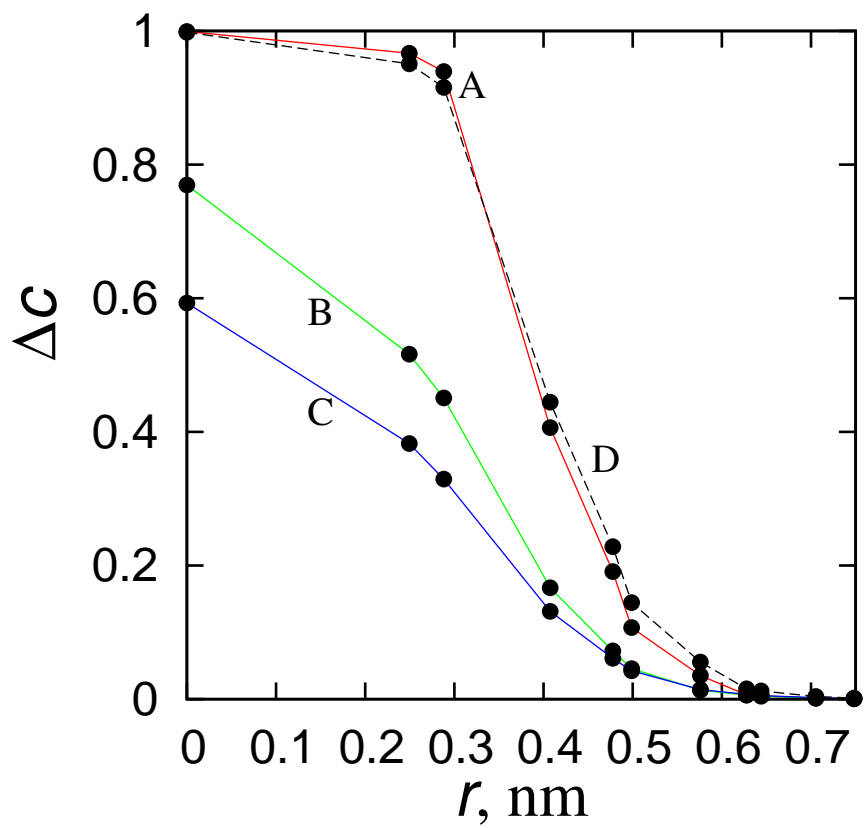
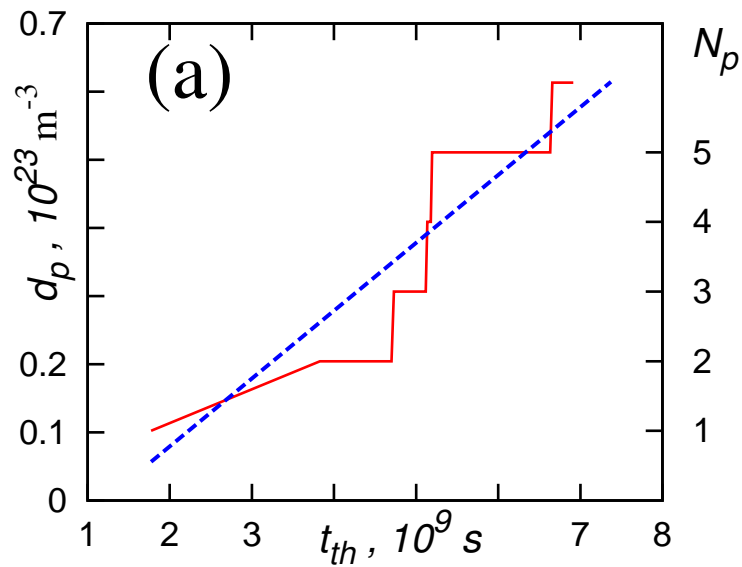
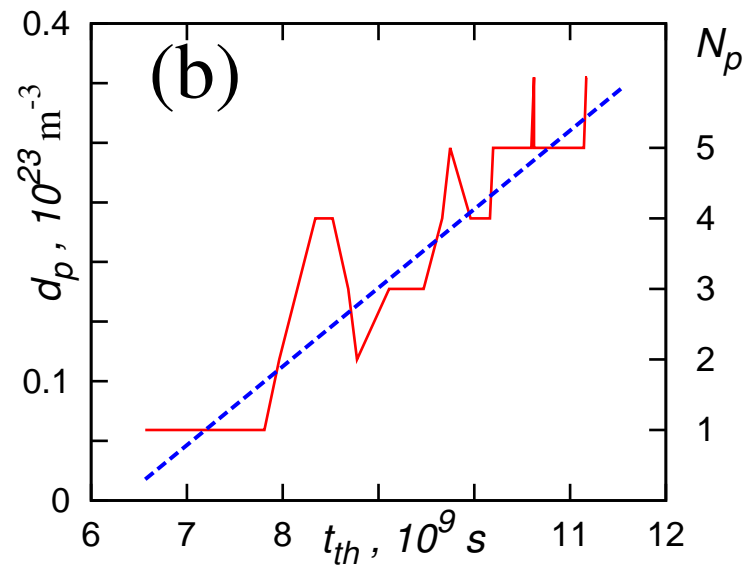
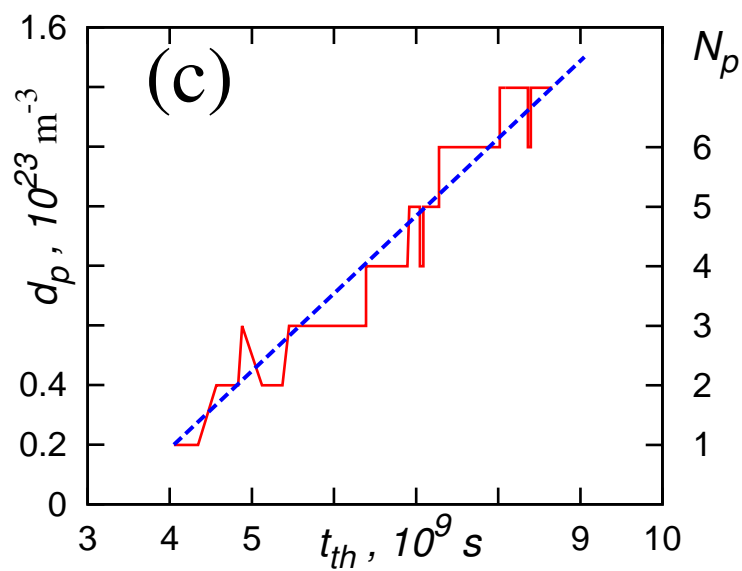
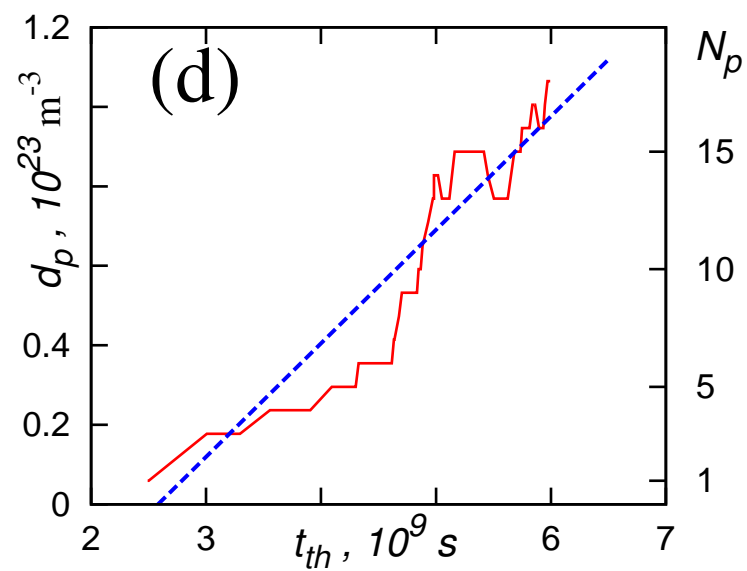
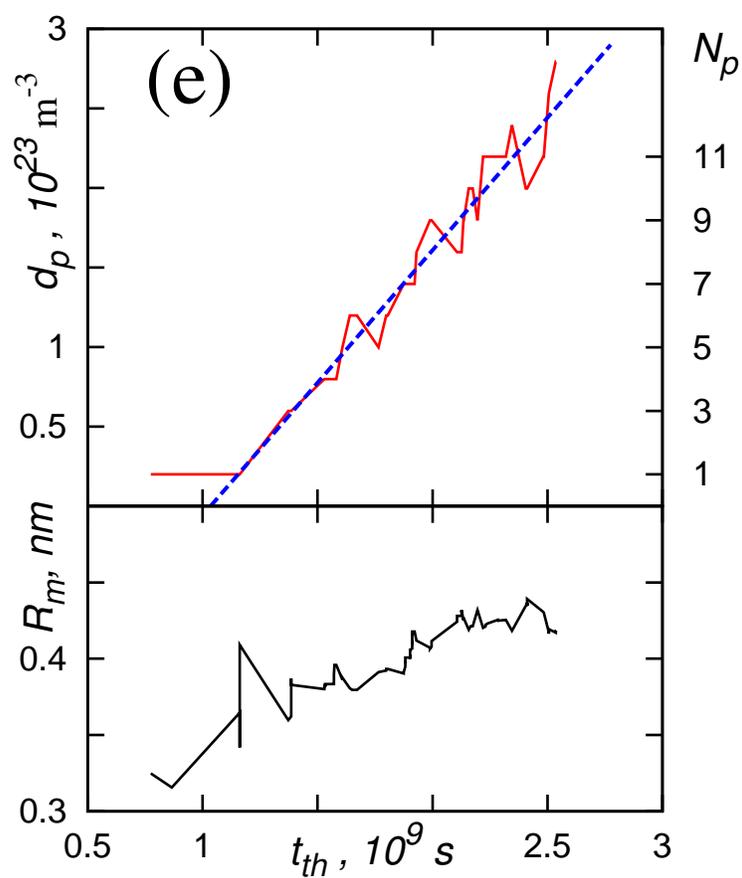
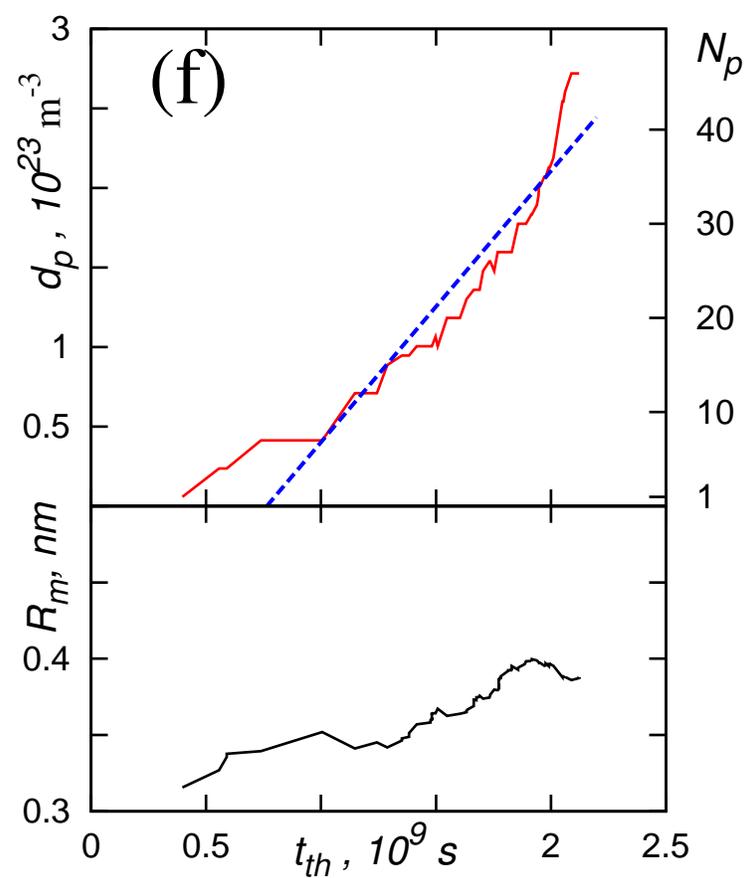
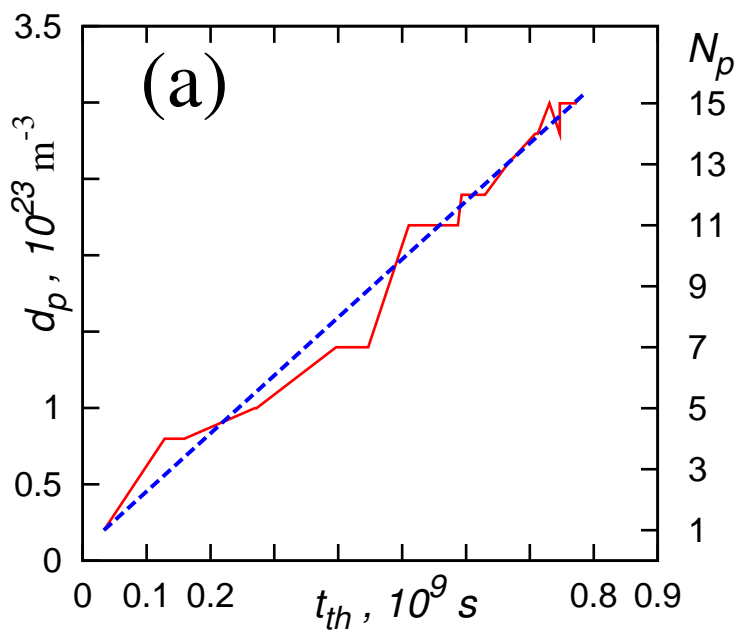
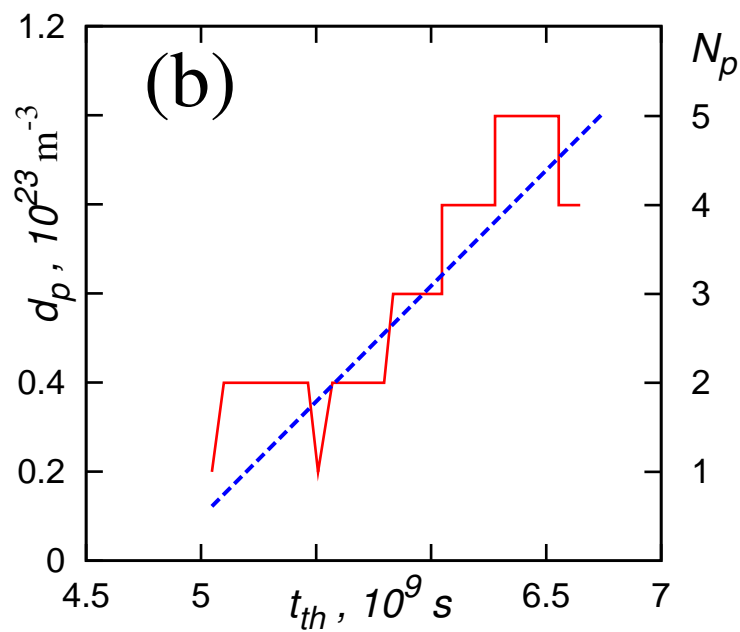
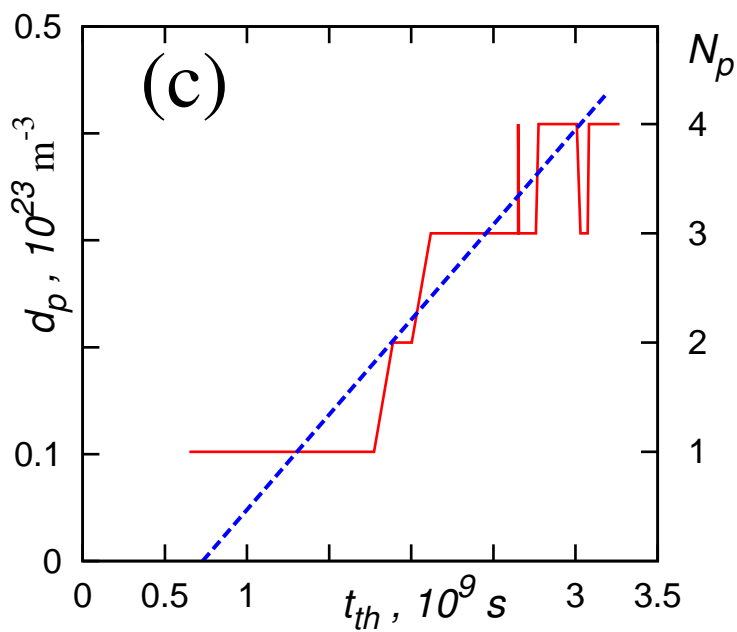
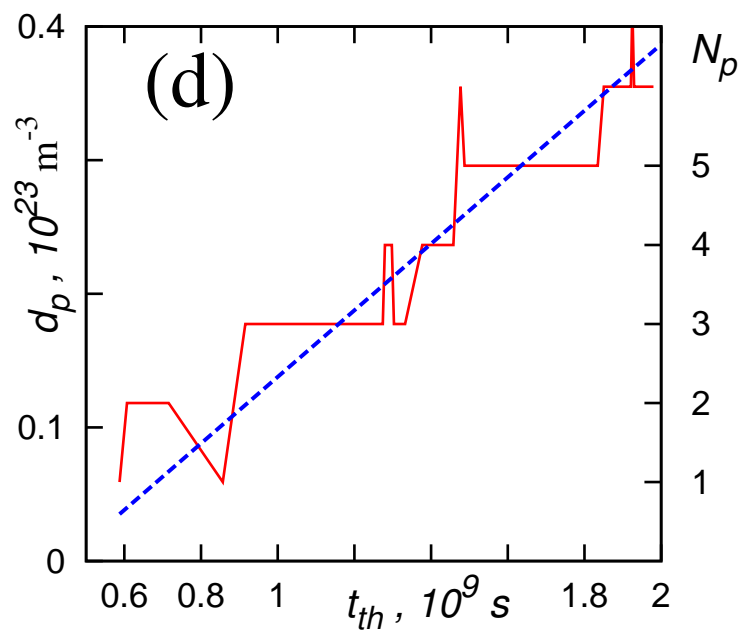
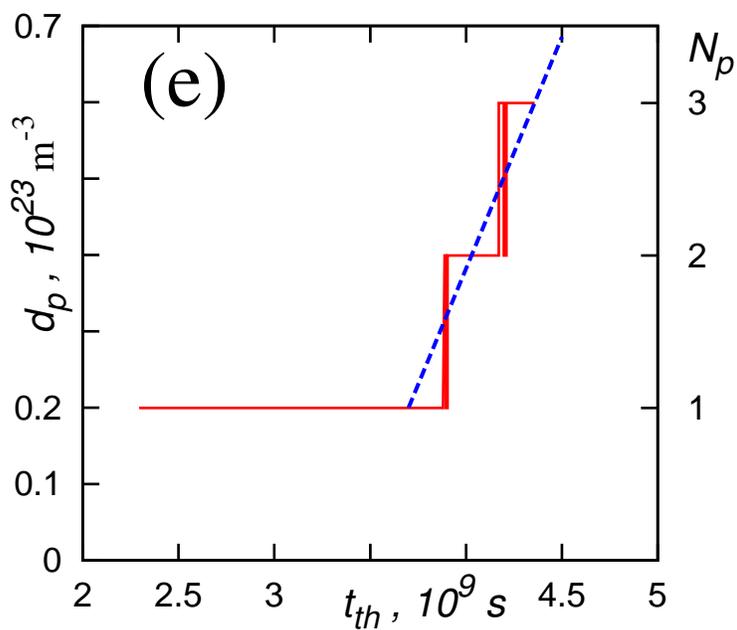
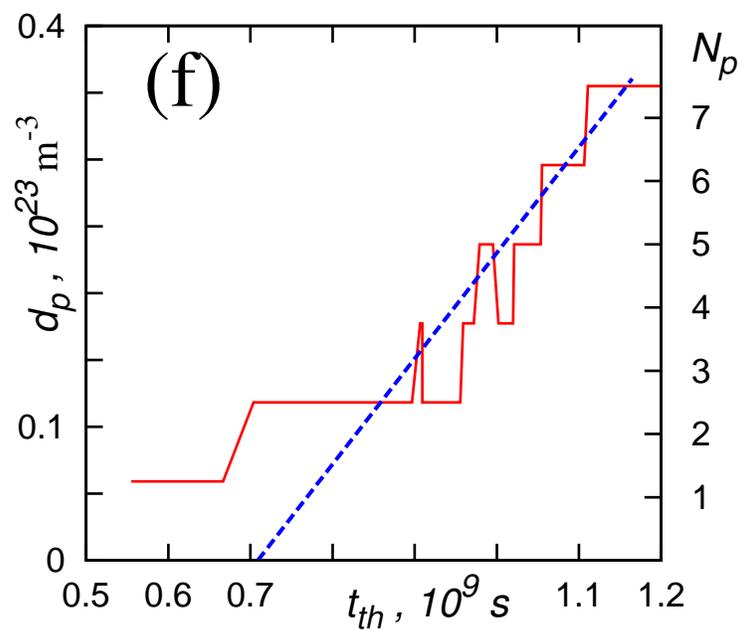
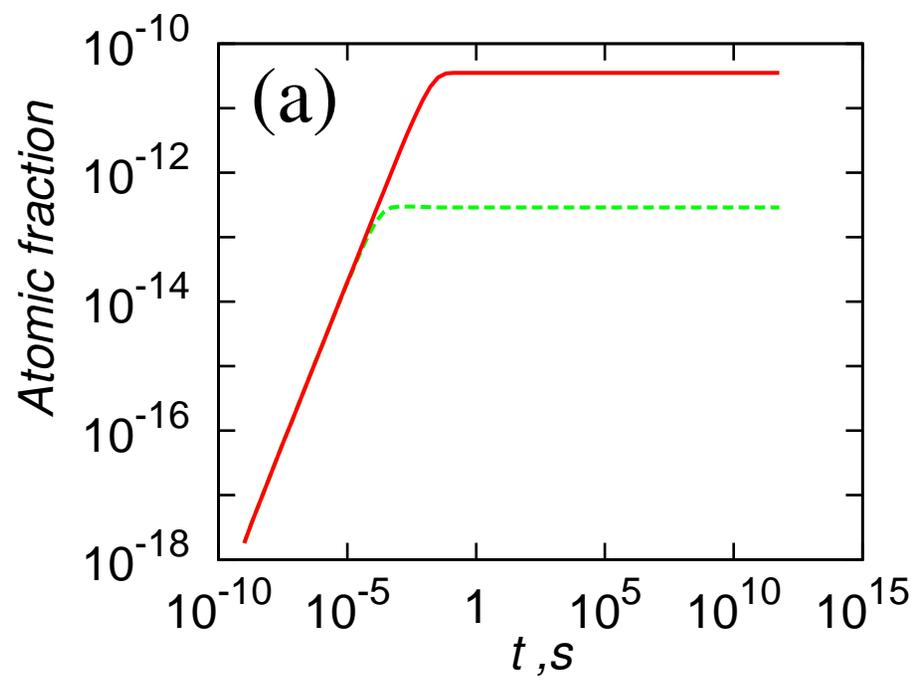
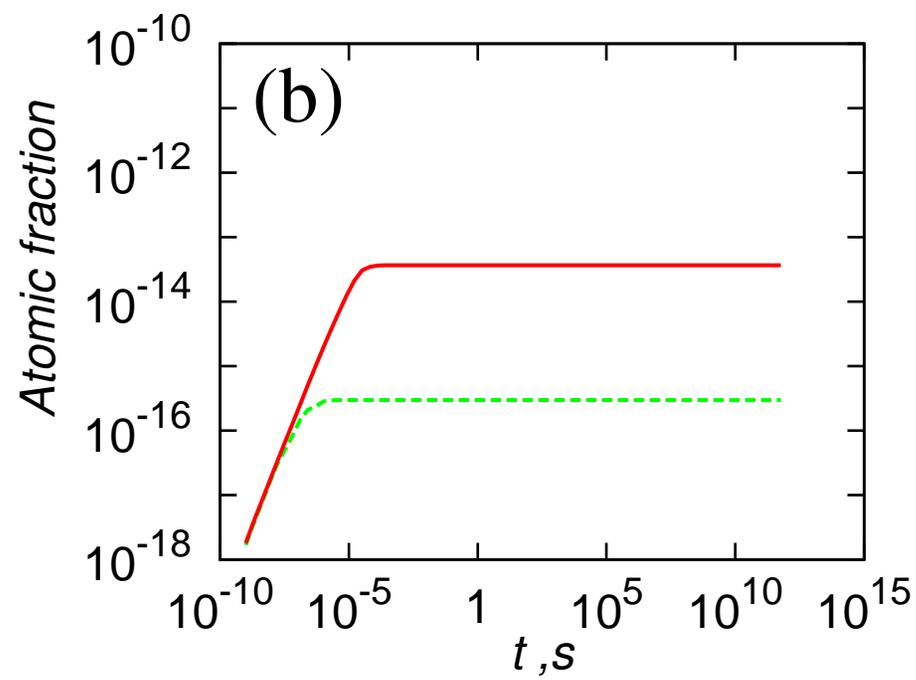
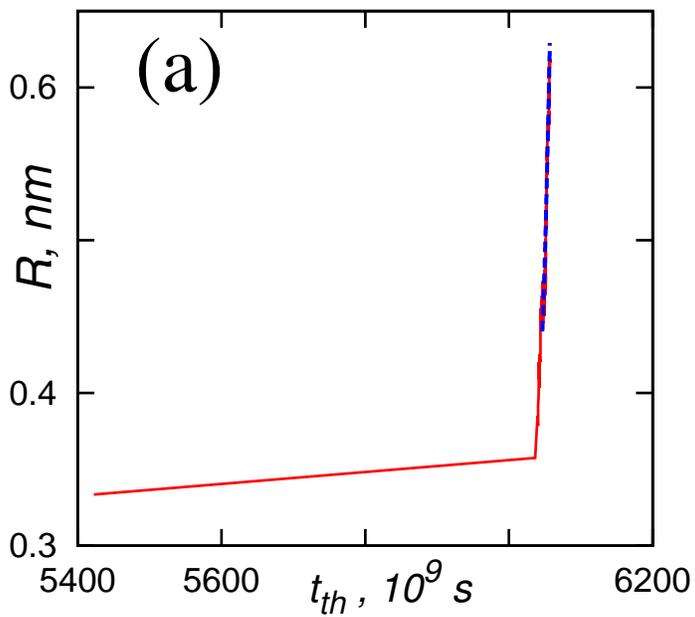
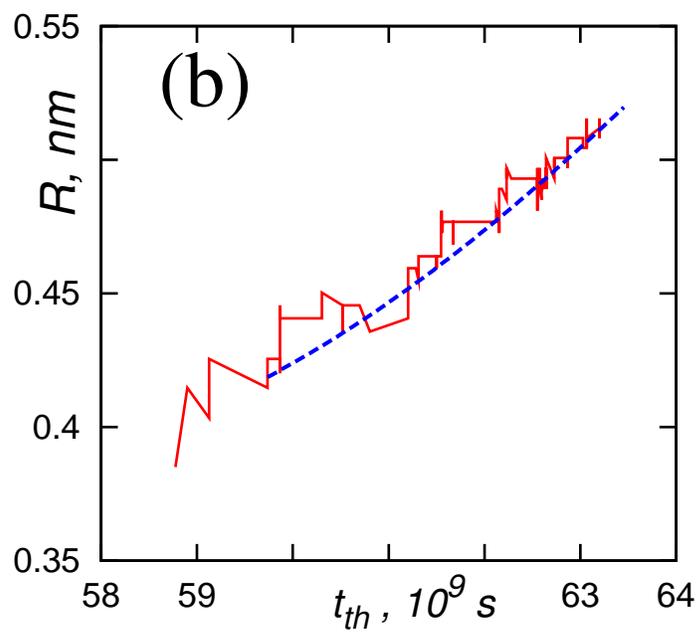
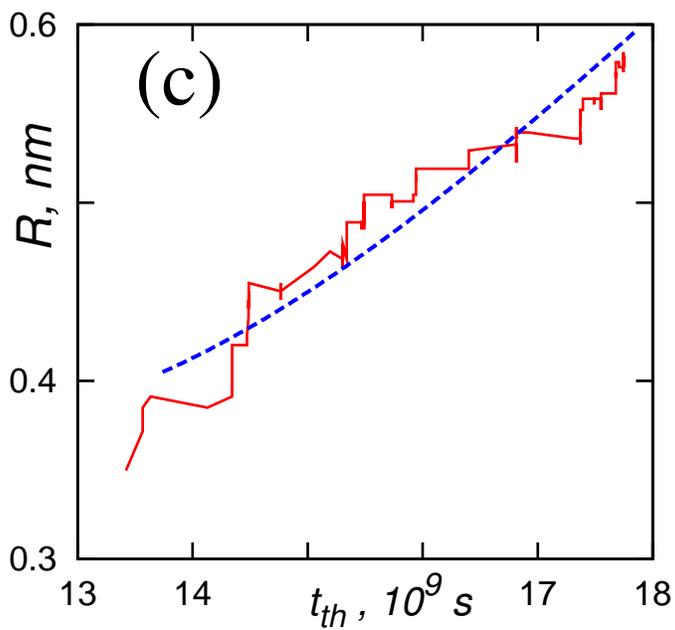
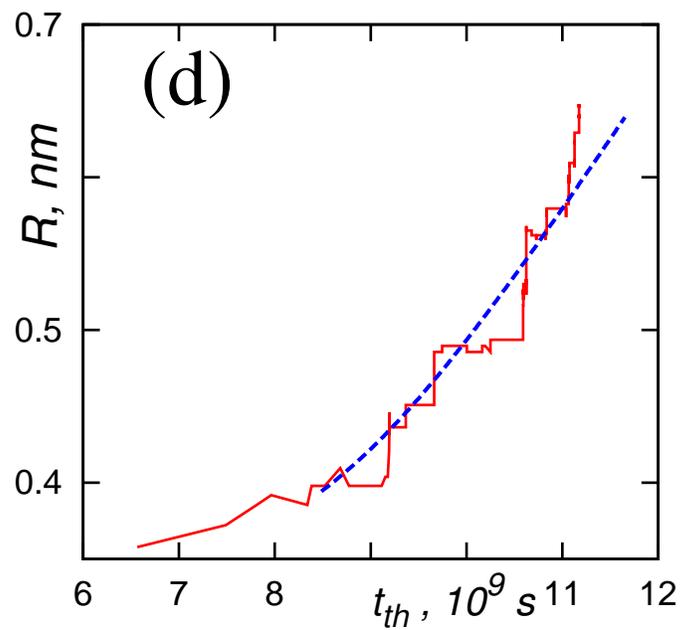
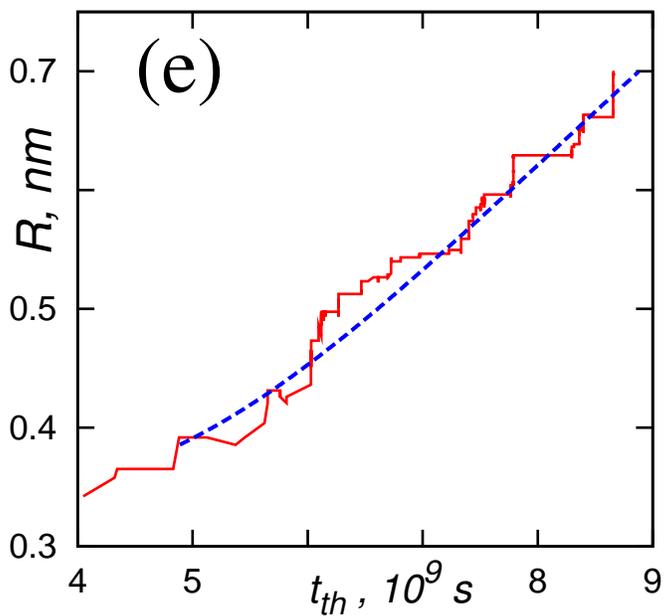
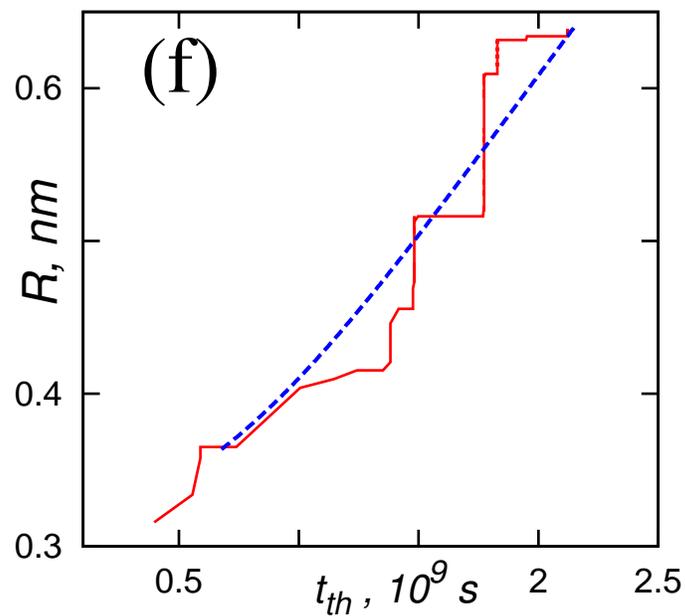
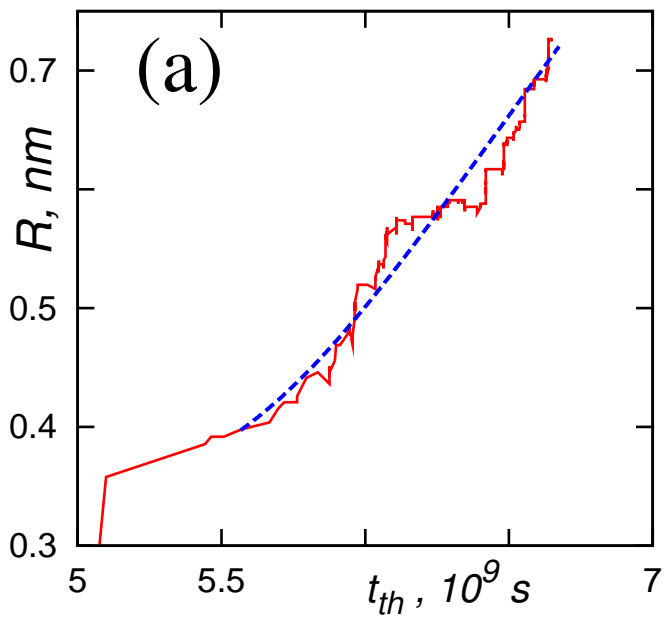
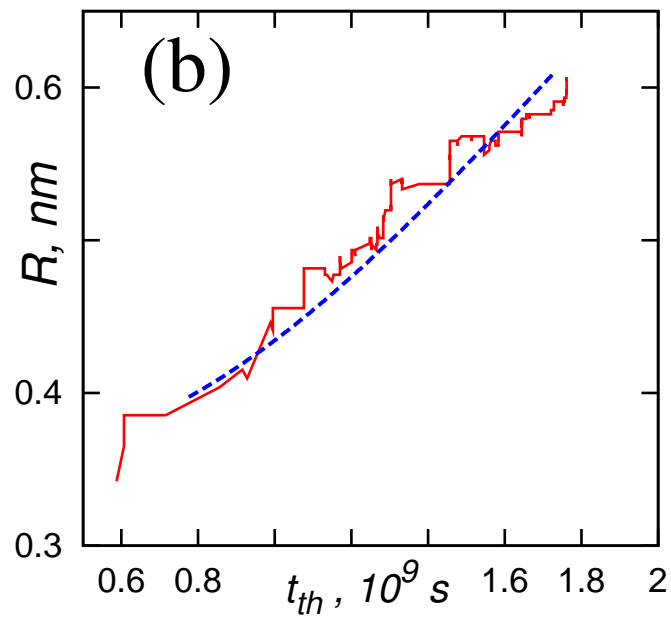
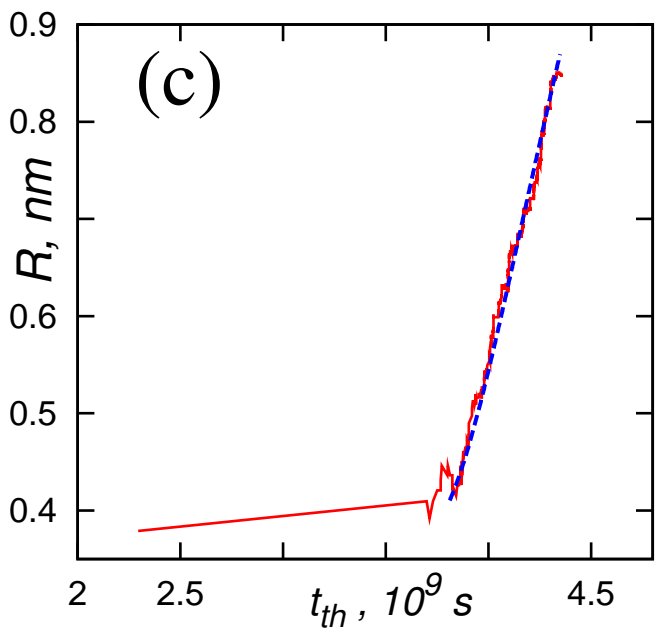
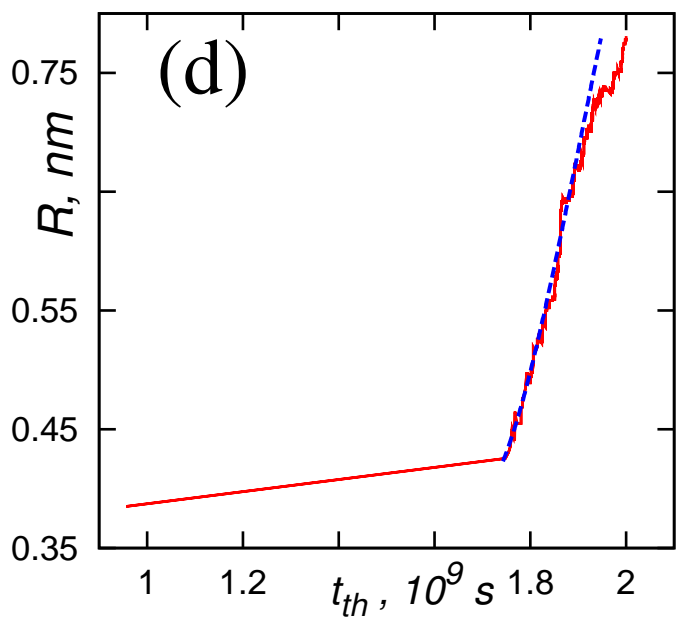
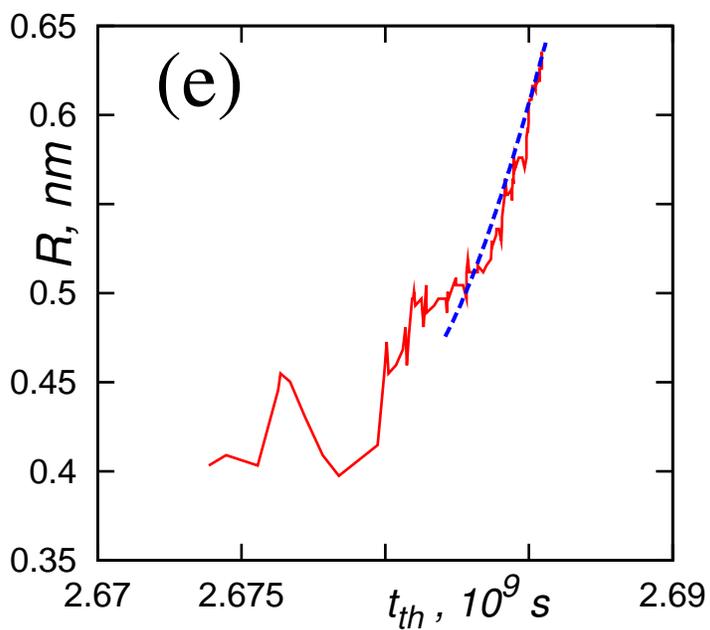
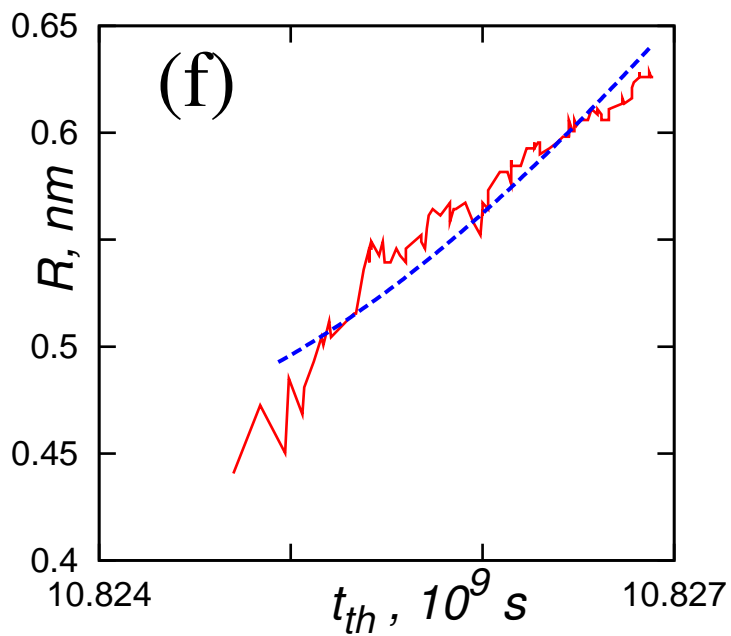
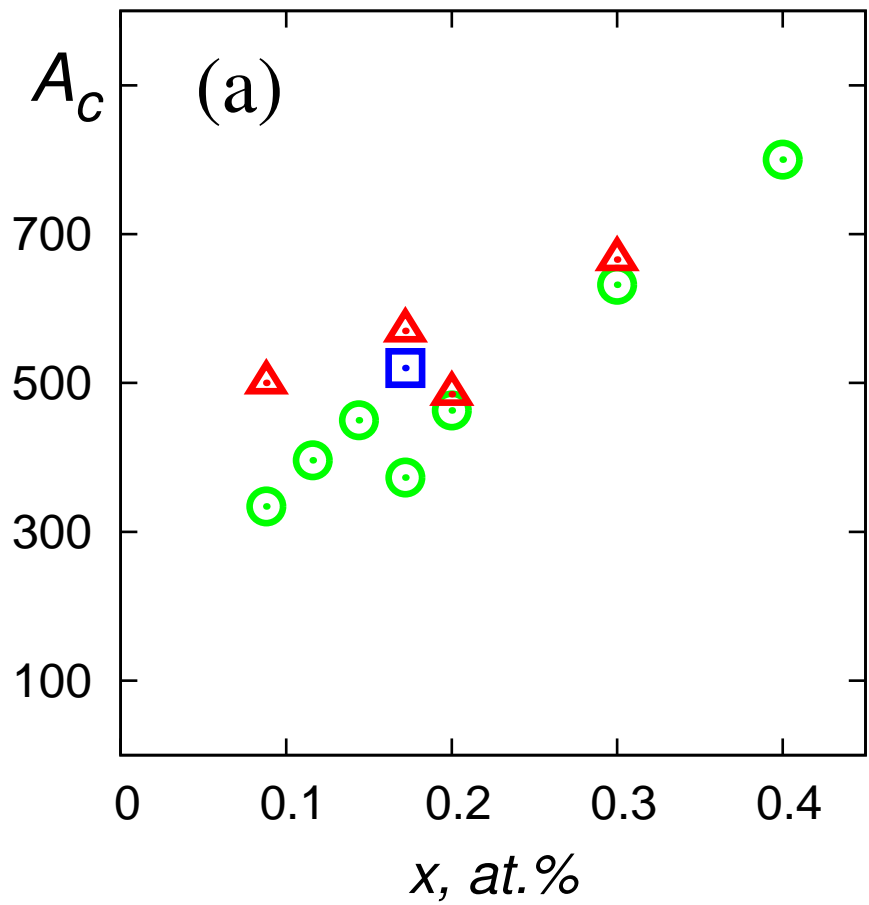
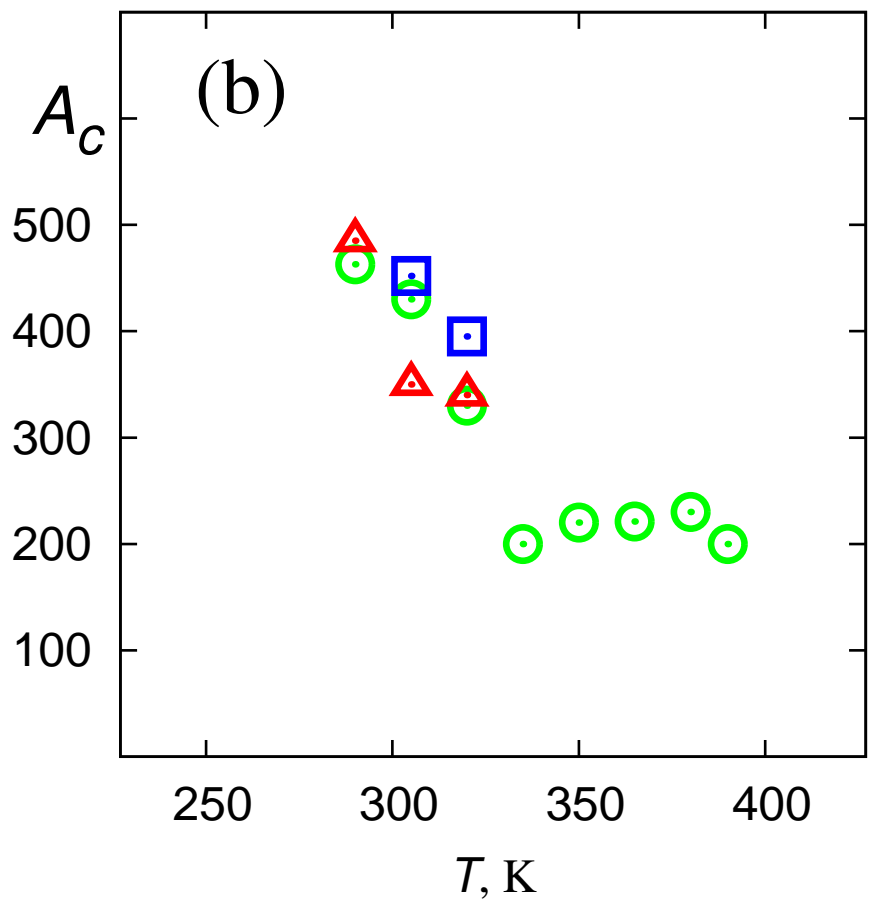
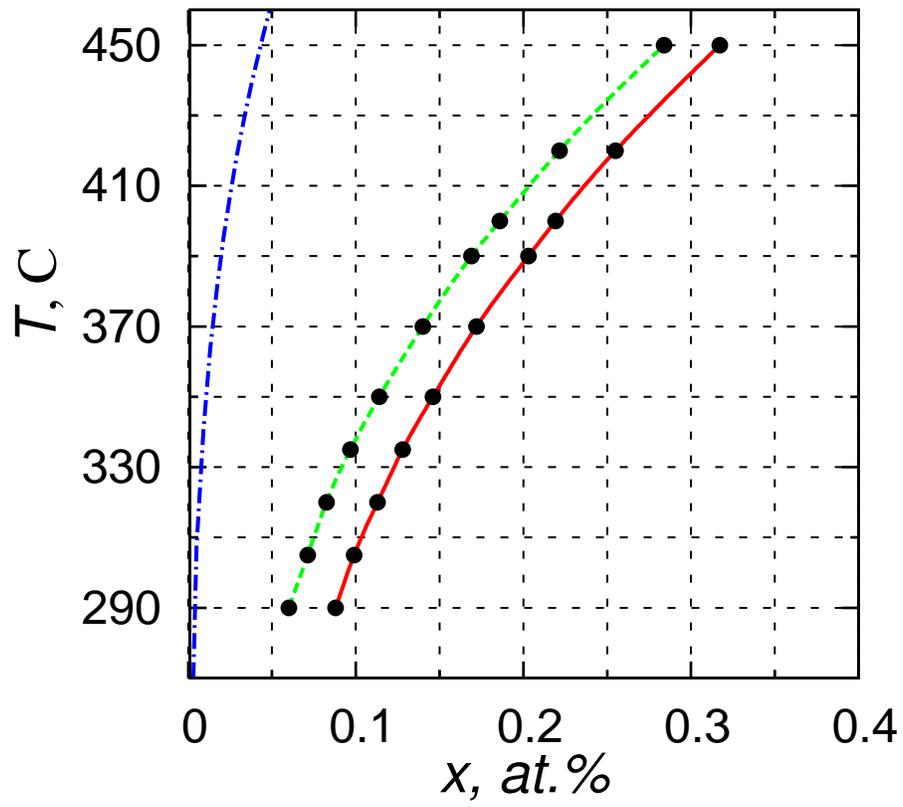
\end{document}